\documentclass[floatfix, notitlepage, nofootinbib, superscriptaddress, longbibliography]{revtex4-1}
\pdfoutput=1

\usepackage[utf8]{inputenc}
\usepackage[T1]{fontenc}
\usepackage{lmodern} 
\usepackage[ngerman,english]{babel}
\usepackage{graphicx} 
\usepackage{microtype} 
\usepackage{mathtools} 
\usepackage{bbold} 
\usepackage{braket} 
\usepackage[shortlabels]{enumitem} 

\usepackage[dvipsnames]{xcolor}

\usepackage{hyperref}
\hypersetup{colorlinks=true,urlcolor=blue,linkcolor=red,citecolor=green!60!black}

\newcommand{\eV}{\,\mathrm{eV}}
\newcommand{\GeV}{\,\mathrm{GeV}}

\newcommand*{\rom}[1]{\text{\expandafter \MakeUppercase{\romannumeral #1}}}

\newcommand{\mpl}{M_{\mathrm{pl}}}
\newcommand{\half}{\frac{1}{2}}
\newcommand{\me}{\mathrm{e}}
\renewcommand{\d}{\mathrm{d}}

\begin{document}


\title{{\LARGE{Misalignment \& Co.}}\\
--\\
{\Large (Pseudo-)scalar and vector dark matter with curvature couplings}
}


\def\andname{\hspace{-1ex}}
\author{Gonzalo Alonso-Álvarez}\email{alonso@thphys.uni-heidelberg.de}
\affiliation{Institut für theoretische Physik, Universität Heidelberg,
Philosophenweg 16, 69120 Heidelberg, Germany}

\author{Thomas Hugle}\email{thomas.hugle@mpi-hd.mpg.de}
\affiliation{Max-Planck-Institut für Kernphysik,
Saupfercheckweg 1, 69117 Heidelberg, Germany}

\author{Joerg Jaeckel}\email{jjaeckel@thphys.uni-heidelberg.de}
\affiliation{Institut für theoretische Physik, Universität Heidelberg,
Philosophenweg 16, 69120 Heidelberg, Germany}


\begin{abstract}
\noindent
Motivated by their potential role as dark matter, we study the cosmological evolution of light scalar and vector fields non-minimally coupled to gravity.
Our focus is on a situation where the dominant contribution to the energy density arises from the misalignment mechanism.
In addition, we also discuss the possibility that dark matter is generated in a stochastic scenario or from inflationary fluctuations.
Even small deviations in the non-minimal couplings from the standard scenarios lead to significant qualitative and quantitative changes. This is due to the curvature-coupling driven superhorizon evolution of the homogeneous field and the non-zero momentum modes during inflation.
Both the relic density yield and the large-scale density fluctuations are affected.
For the misalignment mechanism, this results in a weakening of the isocurvature constraints and opens up new viable regions of parameter space.
\end{abstract}


\maketitle


\section{Introduction}
\label{sec:intro}
Light scalars or vectors very weakly coupled to the Standard Model are phenomenologically and experimentally interesting dark matter candidates~\cite{Preskill:1982cy,Abbott:1982af,Dine:1982ah,Sikivie:2006ni,Nelson:2011sf,Arias:2012az,Jaeckel:2013uva} (cf. also~\cite{Marsh:2015xka} for a recent review). Over recent years a sizeable experimental program has developed to search for their signatures~\cite{Sikivie:1983ip,Horns:2012jf,Budker:2013hfa,Jaeckel:2013sqa,Chung:2016ysi,Graham:2015ifn,Kahn:2016aff,TheMADMAXWorkingGroup:2016hpc,Alesini:2017ifp,Melcon:2018dba,Melcon:2018dba,Du:2018uak,Carney:2019pza}\footnote{While not searching directly for dark matter, experiments like ALPS~\cite{Ehret:2007cm,Ehret:2010mh,Bahre:2013ywa}, CAST~\cite{Zioutas:2004hi,Anastassopoulos:2017ftl} and IAXO~\cite{Irastorza:2011gs,Armengaud:2014gea,Armengaud:2019uso} are searching for suitable light candidates (see also~\cite{Jaeckel:2010ni,Graham:2015ouw,Beacham:2019nyx,Alemany:2019vsk}).} (see~\cite{Jaeckel:2010ni,Hewett:2012ns,Essig:2013lka,Graham:2015ouw,Irastorza:2018dyq} for some overviews).

Their low mass and very small couplings make them automatically stable even on cosmological time scales, without the need to completely forbid their decay by an additional symmetry. Furthermore, they are naturally produced via the misalignment mechanism~\cite{Preskill:1982cy,Abbott:1982af,Dine:1982ah,Nelson:2011sf,Arias:2012az,Jaeckel:2013uva} and as such they generally contribute to at least a fraction of the observed dark matter density.  The initial (homogeneous) field value in our observable Universe can simply correspond to the initial misalignment of the field away from the vacuum, or it can arise ``stochastically''~\cite{Starobinsky:1994bd,Peebles:1999fz,Graham:2018jyp,Guth:2018hsa,Ho:2019ayl,Tenkanen:2019aij} from the accumulated effect of fluctuations during a long phase of inflation. Additional contributions may arise from production via inflationary fluctuations~\cite{Graham:2015rva,Nurmi:2015ema,Kainulainen:2016vzv,Bertolami:2016ywc,Cosme:2018nly,Alonso-Alvarez:2018tus} as well as from decays of precursor particles~\cite{Agrawal:2018vin,Dror:2018pdh,Co:2017mop,Co:2018lka,Bastero-Gil:2018uel,Long:2019lwl}. Modifications to the usual misalignment mechanism have been for example explored in~\cite{Co:2018phi, Co:2018mho}.

So far, most studies of the misalignment mechanism for (pseudo-)scalars have concentrated on the case of minimal coupling to gravity~\cite{Preskill:1982cy,Abbott:1982af,Dine:1982ah}. For vectors, this case typically only yields a small amount of dark matter in today's Universe. In order to enhance the relic abundance, an option is to introduce a curvature coupling which has to be set close to a specific value that makes the evolution equivalent to that of a scalar~\cite{Golovnev:2008cf,Arias:2012az}.
In this paper, we broaden this perspective by considering more general couplings of (pseudo-)scalar and vector fields to the Ricci scalar.
Our main aim is to study their effects on the misalignment mechanism, but we also consider the impact of the curvature couplings on the stochastic scenario~\cite{Graham:2018jyp,Guth:2018hsa} and the contribution of inflationary fluctuations~\cite{Graham:2015rva,Alonso-Alvarez:2018tus} to the dark matter density.

The starting point of our discussion is the usual Einstein-Hilbert action, which is extended to allow for direct couplings of the fields to the Ricci scalar $R$.
In particular, for the scalar case we consider the action
\begin{equation}\label{eq:action_scalar}
S = \int \mathop{\d^4x} \sqrt{-g} \left( \half \left( \mpl^2 - \xi \phi^2\right) R - \half \partial_\mu \phi \partial^\mu \phi - \half m_\phi^2\phi^2 \right),
\end{equation}
where we allow for both positive and negative values of $\xi$. Similarly, the action for the vector field is given by
\begin{equation}\label{eq:action_vector}
S = \int \mathop{\d^4x} \sqrt{-g} \left( \half \left( \mpl^2 + \frac{\kappa}{6} X_\mu X^\mu \right) R - \frac{1}{4}X_{\mu\nu}X^{\mu\nu} - \half m_X^2 X_\mu X^\mu \right).
\end{equation}
Note that the non-minimal coupling in the vector model has the opposite sign with respect to the scalar case. Also, this choice for the normalization of $\kappa$ ensures that the vector field behaves like a minimally coupled scalar for $\kappa = 1$, as we will see in Sec.~\ref{sec:vectors} (such was the convention chosen in~\cite{Arias:2012az, thomasMaster, Dimopoulos:2008yv,Karciauskas:2010as}).

The role of non-minimal couplings to gravity in cosmology has been extensively studied in the literature. The main focus was originally on the construction of inflationary models~\cite{Salopek:1988qh,Hertzberg:2010dc}, which attracted particular interest after it was realized that the Higgs boson can play the role of the inflaton~\cite{Bezrukov:2007ep} if a large value of $\xi\sim 10^4$ is assumed (for a recent review see, e.g.~\cite{Rubio:2018ogq}). The possibility of inflation being driven by a non-minimally coupled vector field was also considered in~\cite{Golovnev:2008cf}. More recently, increasing attention has been payed to the possibility that the dark matter may also enjoy such couplings. The focus of the studies has been on the generation of dark matter during inflation~\cite{Graham:2015rva,Alonso-Alvarez:2018tus,Nurmi:2015ema,Kainulainen:2016vzv,Bertolami:2016ywc,Cosme:2018nly} or at preheating and reheating~\cite{Ford:1986sy,Chung:1998zb,Chung:2001cb,Ema:2015dka,Markkanen:2015xuw,Fairbairn:2018bsw,Ema:2016hlw,Ema:2018ucl,Ema:2019yrd}. Graviton-mediated thermal production has also been explored~\cite{Ema:2016hlw,Ema:2018ucl,Ema:2019yrd,Garny:2015sjg,Tang:2016vch,Tang:2017hvq,Garny:2017kha}.

Aside from phenomenological reasons, there are theoretical motivations to study the couplings delineated above. Generally, these operators need to be considered if general relativity is viewed as an effective field theory~\cite{Donoghue:1994dn}. In addition, the quantization of any scalar field theory in a gravitational background~\cite{Birrell:1982ix} inevitably produces terms like the ones in Eq.~\eqref{eq:action_scalar}. Their presence turns out to be essential for the renormalizability of the energy-momentum tensor in curved backgrounds~\cite{Callan:1970ze}. We will consider ${\mathcal{O}}(1)$ (or smaller) dimensionless couplings, which correspond to interactions of strength $1/M^{2}_\mathrm{pl}$ after the graviton fields are properly normalized. This is the natural size that is a priori expected in a theory of quantum gravity. As an example, such scalar couplings have been found to generically arise~\cite{Narain:2009fy,Narain:2009gb,Percacci:2015wwa,Labus:2015ska,Hamada:2017rvn,Eichhorn:2017sok} in approaches to the quantization of gravity like asymptotic safety~\cite{Weinberg:1976xy} (see~\cite{Eichhorn:2017egq,Bonanno:2017pkg} for recent reviews). 

The situation of vectors is somewhat complicated by gauge invariance and the need to involve a Stueckelberg (or Higgs) mechanism. Constructing the appropriate Stueckelberg terms for the couplings to $R$ leads to a naive violation of unitarity\footnote{We are indebted to Fuminobu Takahashi for discussions on this issue.} already at relatively low energies~\cite{Agrawal:2018vin}. A more detailed investigation, e.g. along the lines of~\cite{Bezrukov:2010jz} would be very interesting but is beyond the scope of this work. Another difficulty for the vector theory is that the sign of the kinetic term of the longitudinal mode becomes negative for a certain range of momenta. This is reminiscent of a scalar particle with a negative kinetic term, i.e. a ghost. As is well known, ghosts are generally dangerous because they destabilize the vacuum of the theory~\cite{Cline:2003gs,Sbisa:2014pzo}. The situation is however more complex for the vector theory at hand. Firstly, the sign of the kinetic term depends on the size of the momentum, and only becomes negative for a finite range of momenta. Secondly, the sign change appears at all only in situations with sufficiently large spacetime curvature. Thirdly and due to the aforementioned sign flip, the kinetic term becomes singular for certain values of the momentum. These problems have been discussed in~\cite{Himmetoglu:2008zp,Himmetoglu:2009qi,Karciauskas:2010as,Nakayama:2019rhg}, but no final conclusion regarding the viability of the theory has been reached. This certainly calls for a more exhaustive investigation, which is however not the aim of the present work. Acknowledging this potential issue, we nevertheless consider the vector scenario and treat the non-minimal coupling as a phenomenological ${\mathcal{O}}(1)$ parameter.

From the point of view of cosmology, the main effect of both scalar and vector non-minimal couplings is that the dark matter fields acquire an additional positive (or negative) mass term during inflation.
As a consequence, the superhorizon evolution becomes fast enough so as to strongly suppress (or enhance) the field value and the fluctuations during the inflationary period, typically in an exponential manner. This has important consequences for the initial field values required to produce today's dark matter density, as even relatively small values of the coupling parameters can lead to drastic shifts in the initial scale that ``naturally'' produces the observed dark matter density.
The fluctuations are affected in a similar way, leading to changes in the isocurvature constraints as well as their contribution to the density today. Both effects will be discussed in detail in this work.

Our paper is structured along the following lines. Section~\ref{sec:scalars}
discusses the effects of non-minimal couplings on the misalignment, stochastic and fluctuation scenarios for (pseudo-)scalars, whereas Section~\ref{sec:vectors} does the same for vectors. For each case, we discuss the amount of dark matter produced as well as constraints from isocurvature fluctuations.
Finally, a short summary and conclusions are given in Section~\ref{sec:conclusions}.

Before continuing let us note that, for simplicity, in the following we will simply write scalars instead of \mbox{(pseudo-)scalars}. That said, all of our results equally apply to pseudoscalar particles.


\section{Scalars}
\label{sec:scalars}


\subsection{Homogeneous scalar field and relic density}\label{sec:scalars_homogeneous_field}
\begin{figure}[t!]
\includegraphics[width=0.75\linewidth]{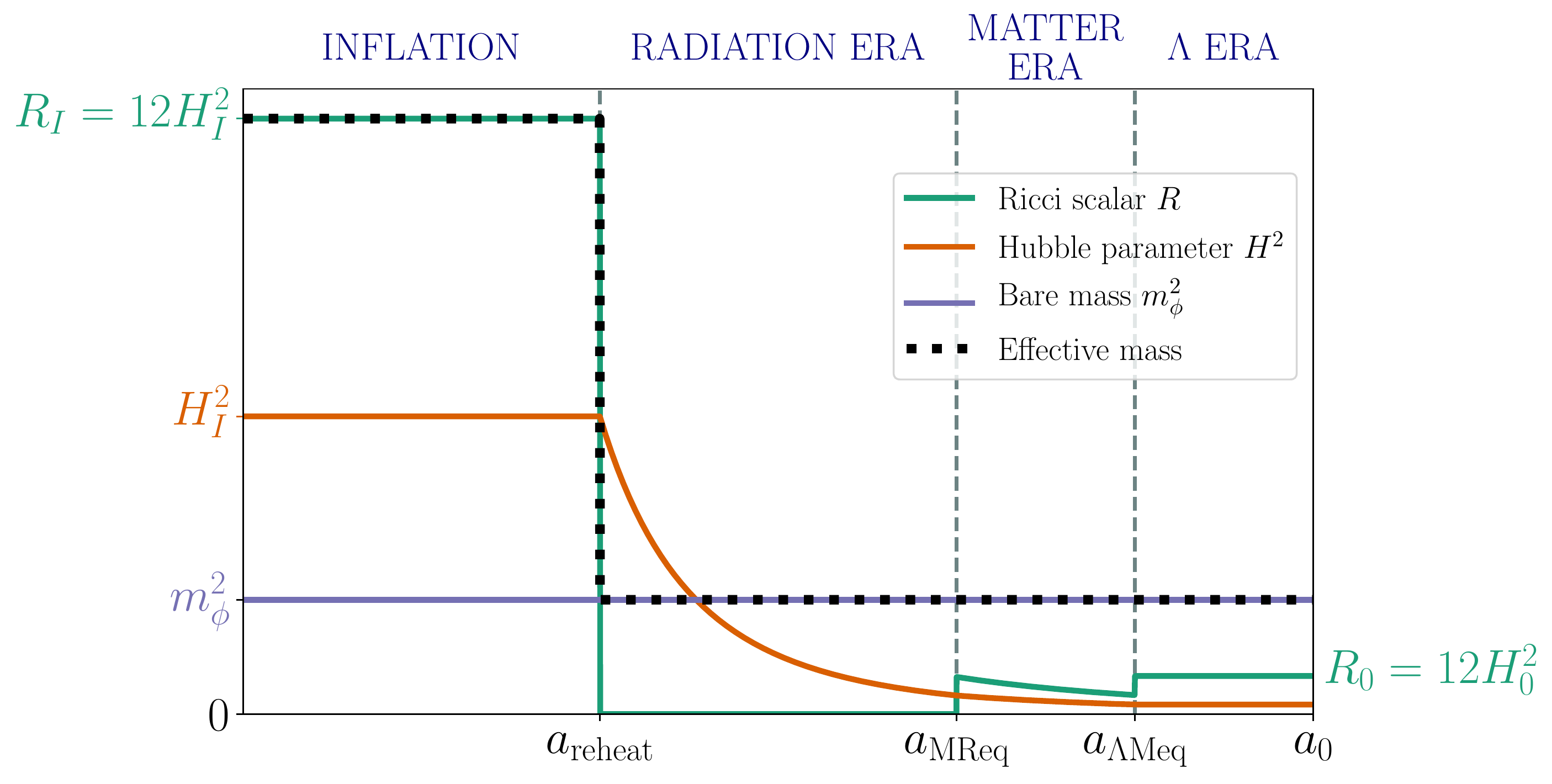}
\caption{\footnotesize{Illustration of the cosmological evolution of the effective mass for the non-minimally coupled scalar (the vector case is analogous). For the purpose of illustration we set $\xi=1$. The field acquires a large effective mass during inflation, when the Ricci scalar is large. As soon as radiation domination starts, $R=0$ and the field acquires its late-time mass given by the explicit mass term $m_\phi^2 \phi^2/2$. By the time of matter-radiation equality, the Hubble parameter is so small that as long as $m_\phi \gtrsim 10^{-28}\eV$, the non-minimal coupling has no significant effect on the effective mass and the evolution any more.} }
\label{fig:effective_mass_evolution}
\end{figure}

The equations of motion (EOMs) for the homogeneous field can be readily derived from the action given in Eq.~\eqref{eq:action_scalar}. Neglecting spatial derivatives, we arrive at
\begin{equation}\label{eq:scalar_eom}
\ddot{\phi} + 3H\dot{\phi} + \left( m_\phi^2 + \xi R \right) \phi = 0.
\end{equation}
As already mentioned in the introduction, the presence of the non-minimal coupling to gravity results in a space-time dependent effective mass for the field. Because of the particular cosmological evolution of the Ricci scalar, which is sketched in Fig.~\ref{fig:effective_mass_evolution}, non-negligible effects will only occur during inflation. This means that the post-inflationary evolution of the scalar field can be described by the standard equations used for the misalignment mechanism (cf.~\cite{Arias:2012az}).
The final relic abundance relative to the observed dark matter one can be expressed as\footnote{Our result differs slightly from the one in~\cite{Arias:2012az} due to the use of the updated value for $\Omega_{\rm DM}$ from~\cite{Akrami:2018odb} and a more careful matching of the initial conditions for the WKB approximation.}
\begin{equation}
\label{eq:scalar_relic_density}
\frac{\Omega_\phi}{\Omega_{\rm DM}} \simeq 5\,\mathcal{F}(T_\star)\left( \frac{\phi_e}{10^{12}\,\mathrm{GeV}} \right)^2 \sqrt{\frac{m_\phi}{\mathrm{eV}}},
\end{equation}
where $\mathcal{F}(T_\star) \equiv (g_\star(T_\star)/3.36)^{3/4} / (g_{\star S}(T_\star)/3.91)$ is an $\mathcal{O}(1)$ function encoding the change in the number of relativistic degrees of freedom between today and the time $t_\star$ when the scalar oscillations start, at $3 H(t_\star) \simeq m_\phi$.
We have explicitly expressed the result in terms of the field value \emph{at the end of inflation} $\phi_e$, to highlight that it is in general different from the initial field value $\phi_s$, which can be identified with the field value \emph{at the start of inflation}. This nontrivial evolution is precisely due to the existence of the $R$-coupling. From now on we assume that the scalar field is responsible for all of the observed dark matter, i.e. $\Omega_\phi = \Omega_\mathrm{DM}$ is enforced if not explicitly stated otherwise.

During inflation\footnote{We approximate inflation by a purely de Sitter expansion with $H=H_I=$ const. Our results do not change qualitatively if we allow for a small non-vanishing $\dot{H}_I$. We will briefly discuss this possibility in Sec.~\ref{sec:conclusions}. We also take reheating to be instantaneous.}, when $R=12\,H_I^2$, the EOM Eq.~\eqref{eq:scalar_eom} is that of a damped oscillator with a constant frequency and can easily be solved.
Due to the non-minimal coupling $\xi \neq 0$, some care has to be put into selecting the appropriate initial conditions. In the minimally coupled case, one can normally assume that $\dot{\phi}_s = 0$ (the subscript $s$ denotes the start of inflation). However, this is not necessarily true in the presence of curvature couplings. Allowing for a non-vanishing $\dot{\phi}_s$ (and assuming that $\phi_s \neq 0$), the general solution is given by
\begin{equation}\label{eq:scalar_homogeneous_field_inflation}
\phi(t) = \phi_s \left( c_1\, \me^{-\half \alpha_- H_I t} + c_2\, \me^{-\half \alpha_+ H_I t} \right),
\end{equation}
where $c_1$ and $c_2$, which are given in Eq.~\eqref{eq:c1_c2}, satisfy $c_{1}+c_{2}=1$ and encode the exact dependence on the initial conditions. Additionally, we have defined
\begin{equation}
\alpha_\pm = 3\pm\sqrt{9-48\xi}.
\end{equation} 
As we will see, all our results only depend on the product $c_1\,\phi_s$. As long as $\dot{\phi_s}\sim H_I\phi_s$ and except for fine-tuned choices of initial conditions, the coefficient $c_1$ is of $\mathcal{O}(1)$. Because of this and for the sake of simplicity, we will henceforth drop it from the equations. For practical purposes, powers of $c_1$ can be easily reinstated by swapping $\phi_s$ for $c_{1}\,\phi_{s}$ at any point in the discussion.
The exact form of the full solution is used for the calculation of the stochastic scenario and can be found in App.~\ref{app:stochastic_scenario}.

For the parameter range of interest, i.e. $\xi<3/16$, $\alpha_\pm$ are always real and positive, and satisfy $\alpha_+>\alpha_-$. Hence, after a long enough time the first term in Eq.~\eqref{eq:scalar_homogeneous_field_inflation} will dominate over the second one and we can approximate the solution by
\begin{equation}
\phi(a) \simeq \phi_s \left( \frac{a}{a_s} \right)^{-\half\alpha_-} = \phi_s\, \me^{-\half\alpha_-N(a)},
\end{equation}
where we use scale factors and number of e-folds instead of time.
For small values of the non-minimal coupling $\xi$, the value of $\alpha_-$ is also small but the product $\alpha_-N(a)$ can be large, resulting in a significant suppression (or enhancement, if $\xi<0$) of the homogeneous field value. In particular, we can relate the value at the end of inflation to the initial condition by
\begin{equation}\label{eq:scalar_homogeneous_field_end_inflation}
\phi_e \simeq \phi_s \me^{-\half\alpha_-N_{\rm tot}}.
\end{equation}
The exponential factor can be enormous depending on the total number of e-folds of inflation $N_{\rm tot}$, which is observationally only bounded from below. Indeed, some models of inflation predict it to be extremely large~\cite{Remmen:2014mia,Graham:2015cka}. In addition to that and in contrast to the minimally coupled case, the field has a non-vanishing time derivative at the end of inflation, 
\begin{equation}\label{eq:phi_e_dot}
    \dot{\phi}_e = -\half \alpha_- H_I \phi_e.
\end{equation}
When $\xi\ll 1$, this is a small effect, as $\d\log\phi_e / \d t \sim\xi$, but it can have important consequences for the post-inflationary evolution of the fluctuations, as we will see in Sec.~\ref{sec:vectors}.

\begin{figure}[t!]
\includegraphics[width=0.6\linewidth]{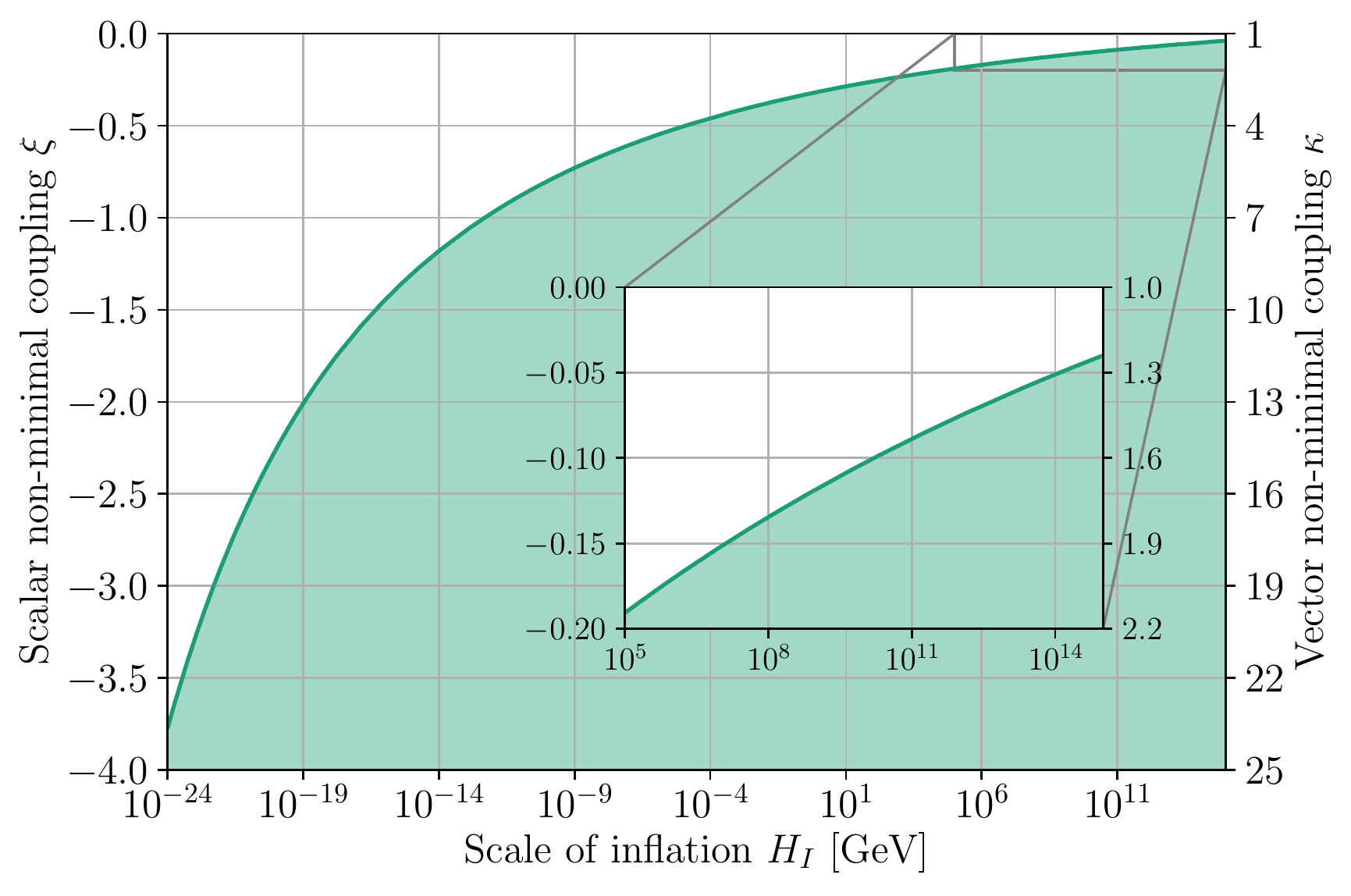}
\caption{\footnotesize{Limit on the size of a negative non-minimal coupling $\xi<0$ (for scalars, left vertical axis) or $\kappa>1$ (for vectors, right vertical axis) as a function of the inflation scale $H_I$, obtained from avoiding a trans-Planckian field excursion (in the absence of extra stabilizing terms in the potential). We make the most conservative assumptions of minimal inflation and initial conditions corresponding to de Sitter vacuum fluctuations, \textit{i.e.} $\phi_s = H_I / 2\pi$ and $\dot{\phi}_s = 0$. The inset zooms into the region of high-scale inflation and small non-minimal couplings.}}
\label{fig:limit_negative_xi}
\end{figure}

As long as there are no additional stabilizing terms in the potential\footnote{For example this could be a self interaction term $\sim \lambda\phi^4$. To avoid trans-Planckian values we would need $\lambda\gtrsim \xi H_{I}^{2}/M^{2}_{\rm pl}$. Note, however, that for $\lambda\phi^2\gtrsim m^2_{\phi}$ the energy density in a homogeneous field dilutes as $\sim a^{-4}$ and the cosmology might be significantly affected (see, e.g.,~\cite{Alonso-Alvarez:2019pfe,Markkanen:2018gcw,Markkanen:2019kpv}). Moreover, stability of the mass under quantum corrections also suggests that self-couplings are severely suppressed, e.g. by a shift symmetry.}, a negative value of $\xi$ induces a runaway potential that can drive $\phi$ into a trans-Planckian field value. Avoiding such a potentially dangerous field excursion sets a limit on the size of the non-minimal coupling. Under the most conservative assumptions of minimal number of inflationary e-folds $N_{\rm tot} = N_{\rm min}(H_I)$ (see App.~\ref{app:length_of_inflation} for an explicit expression for $N_{\rm min}(H_I)$) and initial conditions corresponding to de Sitter vacuum fluctuations $\phi_s=H_I/2\pi$, the limit on $\xi$ only depends on the Hubble scale of inflation $H_I$ and is shown in Fig.~\ref{fig:limit_negative_xi}. This bound is not contingent on the mass of the field or its potential role as dark matter.
The bound gets stronger for a longer period of inflation according to Eq.~\eqref{eq:scalar_homogeneous_field_end_inflation}.
%

\subsection{Scalar fluctuations and isocurvature perturbations}\label{sec:isocurvature_scalar}

The evolution of quantum fluctuations produced during inflation is also modified by the presence of the non-minimal coupling to gravity. This was studied in~\cite{Alonso-Alvarez:2018tus}, to which we refer for details while presenting here only the main results. The time-evolving gravitational background makes the quantum vacuum transition into an excited state, which can be described as a classical stochastic field on smaller and smaller comoving scales as they leave the horizon during inflation. At that time, each independent Fourier mode becomes a Gaussian-distributed random field with power spectrum (valid for $k\neq 0$)
\begin{equation}\label{eq:ps_scalars_horizon_exit}
\mathcal{P}_{\phi}(k,a_k) = \left( \frac{H_I}{2\pi} \right)^2 F(\alpha_-),
\end{equation}
where
\begin{equation}\label{eq:F_function}
F(\alpha_-) \equiv \frac{2^{2-\alpha_-}}{\pi}\,\Gamma^2\!\left( \frac{3-\alpha_-}{2} \right),
\end{equation}
and we denote $a_k=k/H_I$.
At this point, the result only deviates from the usual scale invariant power spectrum of a minimally coupled, massless field by the $\mathcal{O}(1)$ factor $F(\alpha_-)$. This is to be expected, as within the horizon where $k \gg aH_I$ the modes are relativistic even with respect to the $R$-dependent mass. This changes drastically after horizon exit. The modes become non-relativistic due to their large effective mass, and their classical EOM quickly approaches that of the homogeneous field. This means that the field power spectrum is not frozen on superhorizon scales, but rather evolves as
\begin{equation}
\mathcal{P}_\phi(k,a) = \mathcal{P}_\phi(k,a_k) \left( \frac{k}{a\,H_I} \right)^{\alpha_-}.
\end{equation}
By the end of inflation, the power spectrum of an individual mode is suppressed (or enhanced) by
\begin{equation}\label{eq:scalar_fluctuations_end_inflation}
\mathcal{P}_\phi(k,a_e) = \mathcal{P}_\phi(k,a_k) \me^{-\alpha_-N(k)}.
\end{equation}
Here $N(k)$ is the number of e-folds (counting from the end of inflation) at which the mode $k$ exits the horizon. A precise expression for it is given in Eq.~\eqref{eq:number_efolds_k} in App.~\ref{app:length_of_inflation}.
The crucial difference to Eq.~\eqref{eq:scalar_homogeneous_field_end_inflation} is that the evolution of fluctuations only happens while the modes are superhorizon during inflation. As opposed to $N_{\rm tot}$, $N(k)$ can, under some assumptions, be bounded from above using observations, at least for the modes of cosmological interest~\cite{Liddle:2003as}.
\begin{figure}[t!]
\centering
\includegraphics[width=0.6\linewidth]{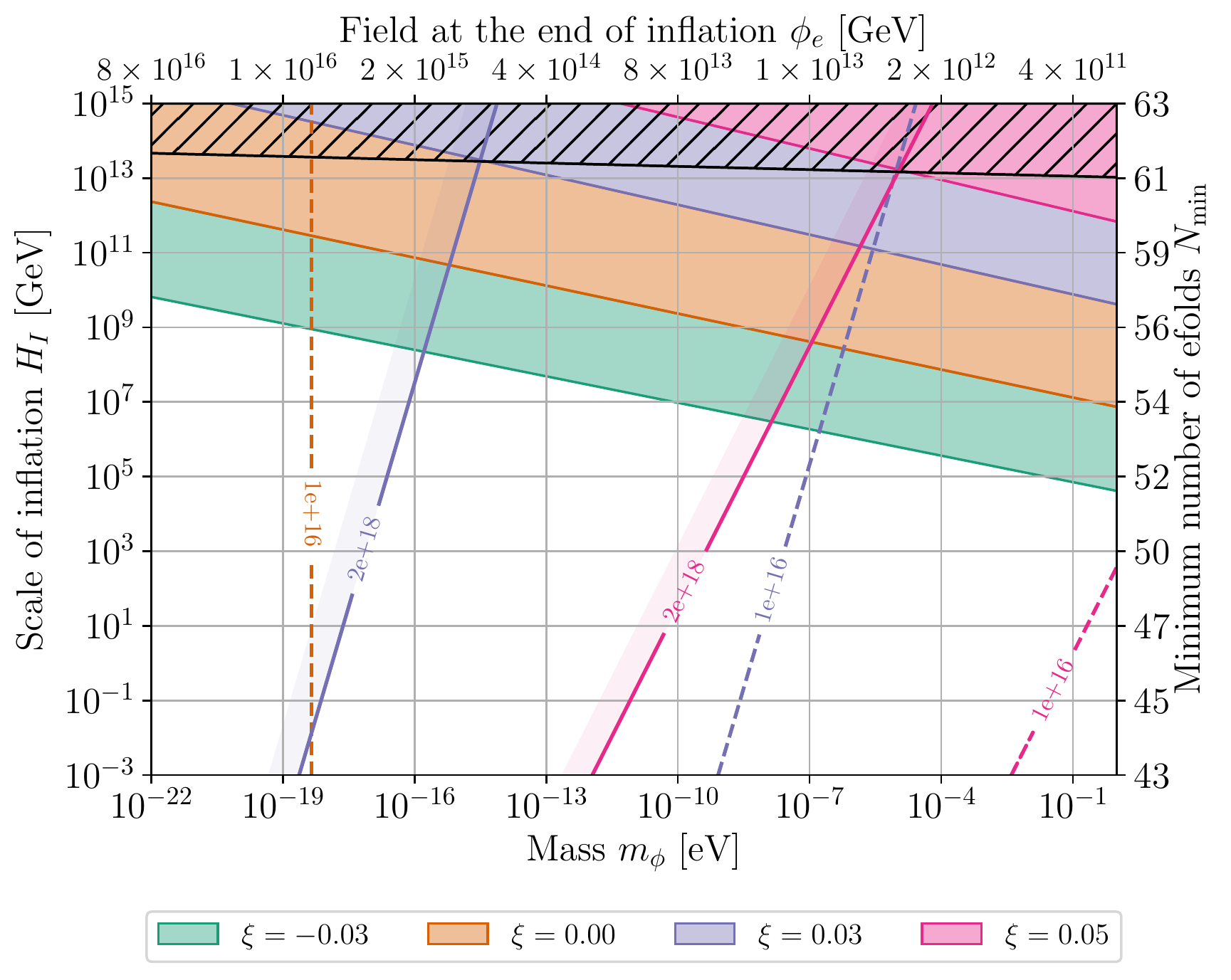}
\caption{\footnotesize{Constraints on the combinations of the inflationary Hubble scale $H_{I}$ and the mass $m_\phi$ of the scalar that allow to produce all of the dark matter via the misalignment mechanism.
The top axis shows the required field value at the end of inflation whereas the right axis shows the minimal number of e-folds of inflation for a given Hubble scale, as given in Eq.~\eqref{eq:N_min}.
The colored regions show the exclusions due to isocurvature constraints for some exemplary values of the non-minimal coupling $\xi$. The lines in the same colors indicate the field value, $\phi_{s}=M_{\rm pl}$ (solid) and $\phi_{s}=10^{16}\,{\rm GeV}$ (dashed) at the beginning of inflation under the assumption of minimal inflation. In the black hatched region there exist no values of $\xi$ and $\phi_s<M_{\rm pl}$ for which the misalignment mechanism generates all the dark matter without producing a too large isocurvature component in the CMB.}}
\label{fig:isocurvature_scalars}
\end{figure}

The field fluctuations translate into density perturbations which become relevant when the dark matter scalar field composes a significant fraction of the energy density of the Universe. Within the misalignment approximation, the fluctuations are always small compared to the homogeneous field value. As the energy density of the coherently oscillating
field is $\rho_\phi\sim \phi^2$, where here $\phi$ denotes the amplitude of the oscillations rather than the instantaneous field value, we can express the density contrast as
\begin{equation}
\delta_\phi\equiv\frac{\delta\rho_\phi}{\rho_\phi} = 2\frac{\delta\phi}{\phi},
\end{equation}
so that its power spectrum can be read off the field one,
\begin{equation}\label{eq:density_field_power_spectrum}
\mathcal{P}_{\delta_\phi}(k) = \frac{4}{\braket{\phi}^2}\mathcal{P}_\phi(k, a).
\end{equation}
Being independent of the inflaton ones, these density fluctuations are generically of isocurvature type and thus subject to constraints arising from their non-observation in the Cosmic Microwave Background (CMB). The evaluation of Eq.~\eqref{eq:density_field_power_spectrum} at the CMB scales is simple\footnote{As we are only interested in the amplitude of the oscillations of the background field and the fluctuations, any potential phase difference between the two is irrelevant.} because the evolution of the scalar field is linear at those large scales, which means that $\mathcal{P}_{\delta_\phi}$ stays constant at all times after horizon exit. Taking advantage of this fact, we evaluate it at the time when the relevant mode $k$ leaves the horizon\footnote{As is commonly done, we use the late time behavior of the fluctuations to evaluate this expression. We estimate that this approximation is valid up to a factor $\lesssim 2$.}. For the CMB modes, this happens $N_{\rm CMB}\equiv N(k_{\rm CMB})$ e-folds before the end of inflation, and thus
\begin{equation}\label{eq:CMB_power_spectrum}
\begin{aligned}
\mathcal{P}_{\delta_\phi}(k_{\rm CMB}) &= \frac{4}{\phi_{s}^2 \,\me^{-\alpha_-(N_{\rm tot}-N_{\rm CMB})}} \left( \frac{H_I}{2\pi} \right)^2 F(\alpha_-) \\
&= \frac{4}{\phi_{e}^2\,\me^{\alpha_-N_{\rm CMB}}} \left( \frac{H_I}{2\pi} \right)^2 F(\alpha_-).
\end{aligned}
\end{equation}
Note that $N_{\mathrm{CMB}} = N_{\mathrm{CMB}}(H_I)$ is determined once the scale of inflation is fixed (see App.~\ref{app:length_of_inflation} for an explicit formula).
We have chosen to display the result both as a function of $\phi_e$ and $\phi_s$ to highlight the difference between both expressions, due to the superhorizon evolution of the modes caused by the non-minimal coupling. 

The non-observation of an isocurvature mode in the CMB by the Planck satellite~\cite{Akrami:2018odb} sets an upper limit\footnote{The isocurvature power spectrum Eq.~\eqref{eq:CMB_power_spectrum} has a nonzero spectral index and is uncorrelated with the adiabatic one. Because of this, we use the limit from the ``axion II" case in the Planck 2018~\cite{Akrami:2018odb} CDI scenario.} on the size of $\mathcal{P}_{\delta_\phi}(k_{\rm CMB})$.
Fig.~\ref{fig:isocurvature_scalars} shows the resulting constraints in the $(m,H_I)$ parameter space for some representative values of $\xi$. A large enough positive non-minimal coupling relaxes the isocurvature bounds, allowing for the misalignment mechanism to produce the observed dark matter density for scales of inflation as high as $\sim 10^{13}\GeV$ irrespective of the mass of the scalar field. This is in stark contrast to the minimally coupled case, where a low enough inflationary scale is required for the misalignment mechanism to be phenomenologically viable. On the contrary and as expected, a negative $\xi$ strengthens the isocurvature constraints.


\subsection{Energy density in fluctuations}
\label{sec:fluctuations_scalar}
In the discussion above we have focused on the production of non-minimally coupled scalar dark matter through the misalignment mechanism. However, as was discussed in~\cite{Alonso-Alvarez:2018tus}, inflationary fluctuations of a non-minimally coupled scalar field can also carry a significant energy density. We now briefly review this alternative production mechanism in order to understand the interplay between the misalignment and the fluctuation-based contribution to the dark matter density.

\begin{figure}[t!]
\centering
\includegraphics[width=0.6\linewidth]{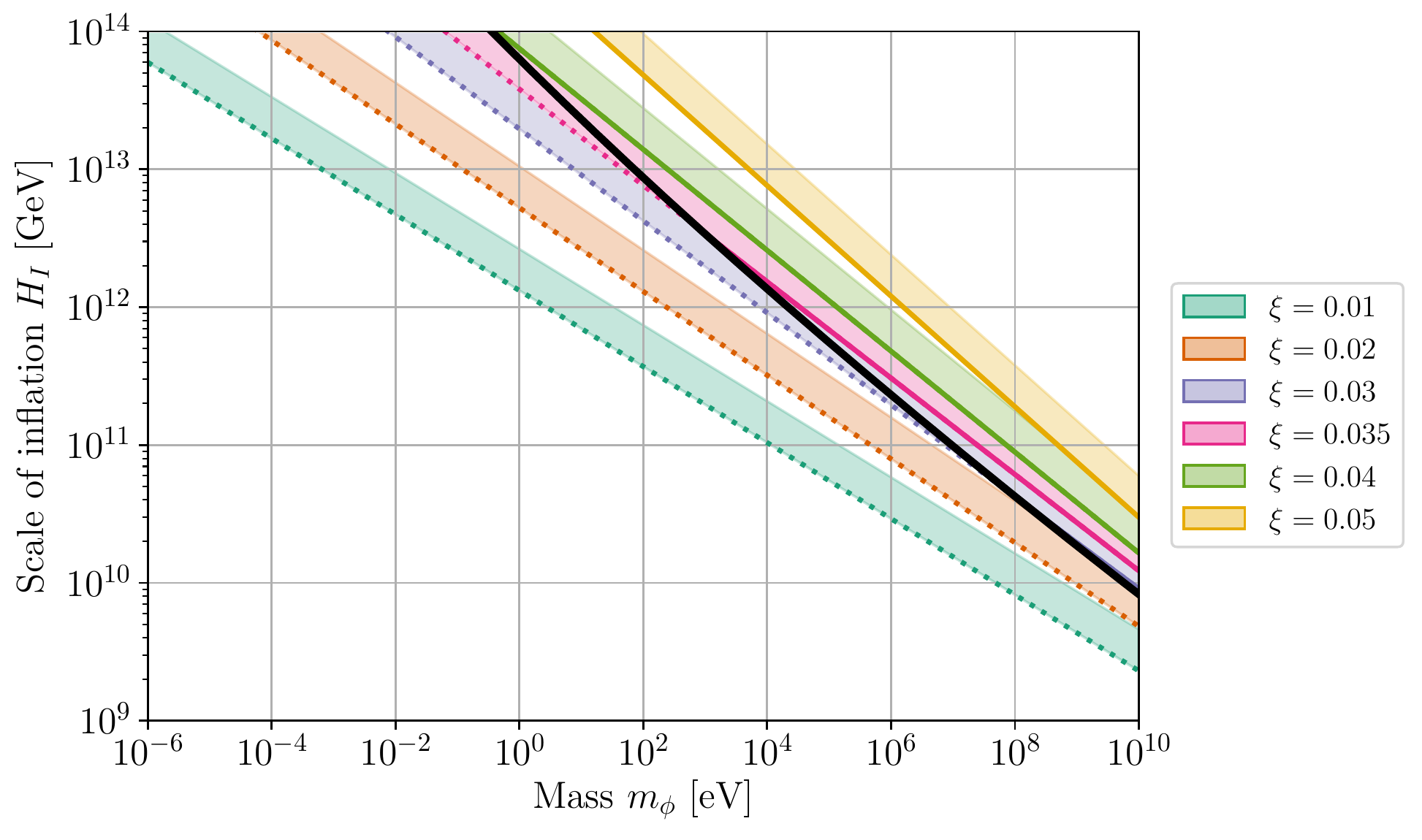}
\caption{\footnotesize{Regions of parameter space where the energy density stored in scalar field fluctuations of inflationary origin makes a sizeable contribution to dark matter. For different choices of the non-minimal coupling $\xi$, the solid and dotted lines correspond to $\Omega_{\phi}=\Omega_{\rm DM}$, while in the partially shaded regions to the right (and beyond) dark matter is overproduced. In addition, we show the constraints arising from the non-observation of isocurvature fluctuations at the CMB scales by the Planck satellite~\cite{Akrami:2018odb}, the regions to the left of the solid black line (corresponding to the dotted lines) are excluded. Note that this figure assumes no contribution to the relic density from the misalignment mechanism.} }
\label{fig:fluctuations_density_isocurvature_scalars} 
\end{figure}

Let us assume that the background homogeneous field value is smaller than the typical fluctuation, in other words, we neglect the energy density generated by misalignment. This is the case if the initial misalignment is too small, or if it gets diluted away by a long period of inflation. To be more precise, any possible initial misalignment is rendered insufficient for dark matter production (cf. Eqs.~\eqref{eq:scalar_relic_density} and~\eqref{eq:scalar_homogeneous_field_end_inflation}) if inflation lasts longer than
\begin{equation}
N_{\rm tot} \gtrsim \frac{1}{\alpha_-}\left( 30 + \half \log{\frac{m_\phi}{\rm eV}}\right),
\end{equation}
where we have taken the extreme initial condition $\phi_s\sim\mpl$. Looking at Fig.~\ref{fig:isocurvature_scalars}, we see that for non-minimal couplings in the range of interest, $\xi\sim 0.01$, a number of efolds in the $10^2-10^3$ range is sufficient to effectively dilute any initial homogeneous field value. In this situation and following~\cite{Alonso-Alvarez:2018tus}, the energy density stored in the inflationary-generated fluctuations of a non-minimally coupled scalar field is
\begin{equation}\label{eq:dark_matter_abundance_fluctuations}
\frac{\Omega_\phi}{\Omega_{\mathrm{DM}}} \simeq \frac{3^{\frac{1}{2}(1+\alpha_-)}}{2^{\frac{11}{4}}\pi^2}\, \frac{F(\alpha_-)}{\alpha_-\left( 1- \alpha_- \right)} \frac{\left[\mathcal{F}(T_\star)\right]^{1+\frac{1}{3}\alpha_-}}{\left[\mathcal{F}(T_\mathrm{r})\right]^{\frac{1}{3}\alpha_-}} \frac{1}{\mpl^2}\, H_{\mathrm{eq}}^{-\frac{1}{2}}\ H_\mathrm{I}^{\half (4-\alpha_-)}\ m^{\half (\alpha_-+1)},
\end{equation}
where $T_r$ is the reheating temperature, $H_{\rm eq}$ denotes the Hubble scale at the time of matter-radiation equality and the rest of notation is as described in previous sections. Fig.~\ref{fig:fluctuations_density_isocurvature_scalars} shows the regions of parameter space where this contribution is sizeable, for different values of $\xi$. It is important to note that the power spectrum of density contrast fluctuations is peaked at the comoving scale
\begin{equation}
    k_\star^{-1}\simeq 1.3\cdot 10^{-6}\,\mathrm{pc}\,\sqrt{\frac{\mathrm{eV}}{m_\phi}},
\end{equation}
which means that the energy density is mainly stored in fluctuations of typical comoving size $k_\star^{-1}$. At larger scales, the power spectrum can be expressed as
\begin{equation}\label{eq:isocurvature_power_spectrum_large_scales}
    \mathcal{P}_{\delta_\phi}(k) \simeq \left( \frac{k}{k_\star} \right)^{2\alpha_-}\mathcal{P}_{\delta_\phi}(k_\star),
\end{equation}
which is valid for $0<\alpha_-<1$ and $k<k_\star$. The amplitude of fluctuations at that scale can be computed as
\begin{equation}
    \mathcal{P}_{\delta_\phi}(k_\star) \simeq \alpha_-^2(1-\alpha_-)^2\cdot I(\alpha_-),
\end{equation}
where $I(\alpha_-)$ is the integral defined as
\begin{equation}
    I(\alpha_-)\equiv \int_{|y-1|<z<y+1} \mathop{\d y}\mathop{\d z} \frac{1}{y^2}\frac{1}{z^2} P(y)P(z),
\end{equation}
with
\begin{equation}
    P(x)\equiv 
    \begin{cases}
    x^{\alpha_-}\,, &x\leq 1\\
    x^{\alpha_--1}\,, &x\geq 1.
    \end{cases}
\end{equation}
The integral is to be evaluated numerically in order to obtain the size of isocurvature fluctuations at CMB scales using Eq.~\eqref{eq:isocurvature_power_spectrum_large_scales}, which can be confronted with the observational constraints by the Planck satellite~\cite{Akrami:2018odb}. As can be seen in Fig.~\ref{fig:fluctuations_density_isocurvature_scalars}, a large enough value of the non-minimal coupling $\xi\gtrsim 0.03$ suppresses the large-scales isocurvature fluctuations below the observational limits. One comment is in order: looking at Eq.~\eqref{eq:isocurvature_power_spectrum_large_scales}, one realizes that the isocurvature power spectrum does not directly depend on the scale of inflation. The reason for this is that both the amplitude of field fluctuations and the variance are quadratic in $H_I$, so the dependence cancels out in the density contrast power spectrum. The apparent dependence on the scale of inflation of Fig.~\ref{fig:fluctuations_density_isocurvature_scalars} is only due to the requirement that the relic energy density matches the observed dark matter abundance.

Compared to the production via the misalignement mechanism, dark matter production from fluctuations is typically viable at much larger masses and higher inflationary Hubble scales.


\subsection{Stochastic scenario}
\label{sec:stochastic_scenario_scalar}

The ``stochastic axion'' scenario was presented in~\cite{Graham:2018jyp,Guth:2018hsa} as a way to generate the dark matter abundance from inflationary fluctuations with wavelengths larger than the current size of the Universe. The accumulation of such fluctuations over a long period of inflation can significantly contribute to the effective homogeneous field in our Hubble patch. This effect can be described as a random walk process of the field value, which receives a ``kick'' every Hubble time from the modes exiting the horizon during inflation. Working against that is the relaxation of the field towards zero, which is driven by the mass of the scalar. For the following discussion, we take a strict definition of the ``stochastic scenario'', requiring that equilibrium is reached to a good approximation. In principle, one could also consider situations where equilibrium is not fully attained. 

This scenario is also significantly impacted by the inclusion of a non-minimal coupling, which acts as a comparatively large, effective mass during inflation. Our calculation follows along the lines of~\cite{Guth:2018hsa} and generalizes it to be valid for scalars with masses comparable to the scale of inflation. This is important as the effective mass $m_\mathrm{eff}^2 \equiv m_\phi^2 + \xi R \approx 12 \xi H_I^2$ is in general not negligible compared to $H_I^2$. This fact also requires us to track the evolution of superhorizon modes during the last $N_{\rm min}$ e-folds of inflation. The details of the calculation are presented in App.~\ref{app:stochastic_scenario}, whereas here we focus on the main results.

Given a long enough inflationary period, the probability distribution of the field value asymptotically approaches a Gaussian distribution with zero mean and variance given by
(see Eq.~\eqref{eq:stochasticfull} for the exact expression),
\begin{equation}\label{eq:stochastic_scalar_field_value}
\left\langle \phi_\delta^2 \right\rangle \simeq \frac{F(\alpha_-)}{\alpha_-} \left(\frac{H_I}{2 \pi}\right)^2  \mathrm{e}^{-\alpha_- N_\mathrm{min}},
\end{equation}
which is valid for positive $\xi$. From this, we deduce that the expected typical value of the field at the end of inflation is of order $\sqrt{\left\langle \phi_\delta^2 \right\rangle}$. The above expression only sums over the modes which are superhorizon today, as those are the ones that contribute to the homogeneous field value in the observable Universe. The first part of Eq.~\eqref{eq:stochastic_scalar_field_value} accounts for the generation and accumulation of fluctuations, while the last factor describes the evolution of the field during the last $N_{\rm min}$ e-folds of inflation. This late time exponential relaxation towards zero is relevant due to the fast evolution driven by the non-minimal coupling to gravity.

In the stochastic scenario, the field ``loses'' memory of the preinflationary initial condition and the observable misalignment angle in each Hubble patch is chosen randomly from the distribution described by Eq.~\eqref{eq:stochastic_scalar_field_value}. Critically, if the effective mass during inflation is large, equilibrium can be reached much faster than in the minimally coupled scenario studied in~\cite{Graham:2018jyp,Guth:2018hsa}. As is shown in App.~\ref{app:stochastic_scenario}, an extra number of e-folds of inflation $\Delta N \equiv N_{\rm tot} - N_{\rm min} > 1/\alpha_-$ on top of the minimum is necessary for equilibrium to be attained. In addition to this, to reach the purely stochastic regime the variance induced by fluctuations has to become larger than the exponentially decaying field value originating from the initial homogeneous $\phi_s$. If we require the induced field value to be a factor of $\gamma$ larger, we get
\begin{equation}\label{eq:stochastic_scalar_N}
\Delta N > \frac{1}{\alpha_-} \ln\left[ \gamma \frac{2^{2 + \alpha_-} \pi^3}{\Gamma^2\!\left( \frac{3 - \alpha_-}{2} \right)} \frac{\alpha_-}{(\alpha_+ - \alpha_-)^2} \left( \frac{\dot{\phi_s}}{H_I^2} + \frac{1}{2} \alpha_+ \frac{\phi_s}{H_I} \right)^2 + 1 \right].
\end{equation}
The two results in Eqs.~\eqref{eq:stochastic_scalar_field_value} and~\eqref{eq:stochastic_scalar_N} generalize the ones found in~\cite{Graham:2018jyp,Guth:2018hsa}, allowing for arbitrary initial conditions and large effective masses.

The stochastic scenario is also prone to generating too much isocurvature fluctuations at the CMB scales. From the point of view of subhorizon modes, the homogeneous field value generated in the stochastic regime is indistinguishable from one originating from a homogeneous initial condition. This implies that Eq.~\eqref{eq:CMB_power_spectrum} also holds in the stochastic case, which allows us to express the isocurvature component as
\begin{equation}
\mathcal{P_{\delta_\phi}}(k_{\rm CMB}) \simeq 4\alpha_-\,\mathrm{e}^{\alpha_- (N_{\rm min} - N_{\rm CMB})}.
\end{equation}
The exponential factor arises due to the slight difference between the largest observable scale and the ones that are accessible at the CMB, which corresponds to about $7$ e-folds (see App.~\ref{app:length_of_inflation}). As long as $\alpha_-$ is small, this is a small correction and the result is dominated by the factor in front. It is easy to understand where this factor comes from: comparing Eq.~\eqref{eq:stochastic_scalar_field_value} with Eq.~\eqref{eq:ps_scalars_horizon_exit}, we see that the accumulated field variance of a large number of superhorizon modes in enhanced by a factor of $1/\alpha_-$ with respect to the amplitude of the fluctuations of an individual mode. Given the Planck~\cite{Akrami:2018odb} constraint on isocurvature fluctuations $\mathcal{P}_I(k_{\rm CMB})\lesssim 10^{-9}$, we conclude that the stochastic scenario can only be realized for very small values of the non-minimal coupling of order $\xi \lesssim 10^{-10}$. This should apply to any model in which the effective mass is non-negligible compared to $H_I$, independently of the origin of $m_\mathrm{eff}$.


\section{Vectors}
\label{sec:vectors}
After having discussed the misalignment, stochastic and fluctuations scenario for a scalar field, let us now turn to the vector case and see how these scenarios can be realized there.

\subsection{Homogeneous vector field and relic density}
Starting from the action Eq.~\eqref{eq:action_vector}, we derive the EOMs for the homogeneous field~\cite{Golovnev:2008cf,Arias:2012az,thomasMaster}
\begin{equation}
\chi_0 = 0 \quad \mathrm{and} \quad \ddot{\chi}_i + 3 H \dot{\chi}_i + \left(m_X^2 + \frac{1 - \kappa}{6} R \right) \chi_i = 0,
\end{equation}
which are given in terms of the physical field $\chi_\mu \equiv X_\mu / a$, so that the energy density can (for an approximately homogeneous field) be written as~\cite{Golovnev:2008cf,Arias:2012az,thomasMaster}
\begin{equation}
    \rho_\chi = \sum_{i=1}^{3}\, \left[ \half \dot{\chi}_i^2 + \half m_X^2 \chi_i^2 + (1-\kappa) \left( \half H^2\chi_i^2 + H \dot{\chi}_i\chi_i \right) \right].
\end{equation}
The advantage of this field redefinition is that the EOM for each of the spatial components coincides with the scalar one, Eq.~\eqref{eq:scalar_eom}, after the identification
\begin{equation}\label{ident}
\frac{(1 - \kappa)}{6} \longleftrightarrow \xi.
\end{equation}
In particular, a non-minimally coupled vector with $\kappa = 1$ behaves as a scalar with a minimal coupling to gravity ($\xi = 0$). Because of this correspondence, the discussion of the misalignment mechanism made in Sec.~\ref{sec:scalars_homogeneous_field} carries over to the vector case, as do all the evolution equations (cf. Eq.~\eqref{eq:scalar_homogeneous_field_inflation} ff.). Let us then simply rewrite the main results using the notation of this section. The relic density of vectors can be expressed as
\begin{equation}
\label{eq:vector_relic_density}
\frac{\Omega_\chi}{\Omega_{\rm DM}} \simeq 5\,\mathcal{F}(T_\star)\left( \frac{\chi_e}{10^{12}\,\mathrm{GeV}} \right)^2 \sqrt{\frac{m_X}{\mathrm{eV}}},
\end{equation}
where we denote $\chi\equiv \left| \Vec{\chi} \right|$.
As in the scalar case, the field value at the end of inflation $\chi_e$ is different than the initial one $\chi_s$ due to the superhorizon evolution caused by the coupling to $R$. Both are related by
\begin{equation}\label{eq:vector_homogeneous_field_end_inflation}
\chi_e \simeq \chi_s \,\me^{-\half \beta_- N_{\mathrm{tot}}},
\end{equation}
with an analogous definition\footnote{The parameter $\beta_-$ defined here is related to $\nu$ used in~\cite{thomasMaster,Dimopoulos:2008yv} by $\beta_- = 3 - 2 \nu$, neglecting terms of $\mathcal{O}(m_X / H_I)$.} to the scalar case,
\begin{equation}
\beta_\pm \equiv 3 \pm \sqrt{1 + 8 \kappa}.
\end{equation}
Note that we are using the conventions described in Sec.~\ref{sec:scalars_homogeneous_field} for the field value at the beginning of inflation.
Depending on whether $\kappa$ is larger or smaller than one, there can be a substantial enhancement or suppression of the homogeneous field value during inflation. As in the scalar case, an absolute limit can be placed on $\kappa>1$, independently of the dark matter assumption, by avoiding a trans-Planckian field excursion. Using the identification of Eq.~\eqref{ident}, this limit is analogous to the one for scalars. Both are shown simultaneously in Fig.~\ref{fig:limit_negative_xi}.


\subsection{Vector fluctuations and isocurvature perturbations}
\label{sec:isocurvature_vector}

Like a scalar field, a vector field present during inflation acquires a spectrum of isocurvature perturbations. We now study the role of the non-minimal coupling in the generation and evolution of such fluctuations.


\subsubsection{Generation during inflation}

For vectors, the situation is similar to the scalar case, albeit a bit more complex due to the multicomponent nature of the field. The best way to deal with the perturbations is to split them into transverse ($\perp$) and longitudinal ($\parallel$) modes and address each polarization separately.

\medskip
\paragraph{Transverse fluctuations.}
The EOMs of the fluctuations in momentum space and for the two transverse polarizations $\delta\chi^\perp_i$ read
\begin{equation}
\ddot{\delta\chi}^\perp_i + 3 H \dot{\delta\chi}^\perp_i + \left( m_X^2 + \frac{1 - \kappa}{6} R + \frac{k^2}{a^2} \right) {\delta\chi}^\perp_i = 0,
\end{equation}
which is identical to the expression in the scalar case (cf.~\cite{Alonso-Alvarez:2018tus}) with $(1-\kappa) / 6 \leftrightarrow \xi$.
We can therefore directly translate the results of Sect.~\ref{sec:isocurvature_scalar}: the power spectrum for non-vanishing momentum modes outside the horizon ($0\neq k\ll a H_I$) is
\begin{equation}\label{eq:ps_transverse_fluctiations_inflation}
\mathcal{P}^\perp_{\chi_i} (k, a) = \left( \frac{H_I}{2 \pi} \right)^2 F(\beta_-) \left( \frac{k}{a H_I} \right)^{\beta_-},
\end{equation}
with $F(\cdot)$ as defined in Eq.~\eqref{eq:F_function}.

\medskip
\paragraph{Longitudinal fluctuations.}
In this case, the discussion is simplified by making use of conformal time $\tau$ and of the field redefinition $f \equiv a^2\,{\delta\chi}^\parallel = a\,{\delta X}^\parallel$. The full EOM for the mode functions then reads
\begin{equation}\label{eq:EOM_longitudinal_fluctuations_inflation}
    f^{\prime\prime} - \frac{2}{\tau}\frac{2\kappa}{k^2\tau^2 - 2\kappa}f^{\prime} + k^2 \left( 1 - \frac{2\kappa}{k^2\tau^2} - \frac{2}{k^2\tau^2 - 2\kappa} \right)f=0
\end{equation}
where we have assumed that $m_X^2\ll H_I^2$ so that the bare mass has been neglected. It can easily be reinstated by substituting $\kappa\rightarrow\kappa - m_X^2/(12H_I^2)$ at any point in the derivation. Although we are not able to solve Eq.~\eqref{eq:EOM_longitudinal_fluctuations_inflation} analytically, we can study its most relevant limits.
\begin{itemize}
\item \underline{Subhorizon limit $|k\tau|\gg 1$}. The mode equation simplifies in this limit to
\begin{equation}
f^{\prime\prime} +  k^2 f = 0.
\end{equation}
We recognize the equation of motion of a harmonic oscillator in Minkowski space: as expected, the mode functions in the deep subhorizon limit can be described as quantum fluctuations of the vacuum. This allows us to set the initial condition for the evolution,
\begin{equation}
f(\tau \rightarrow - \infty) = \frac{1}{\sqrt{2k}} \me^{-i k \tau},
\end{equation}
which is the usual Bunch-Davies vacuum. Note that the units of Fourier transformed fields differ from the ones in position space by a factor of [mass]$^{3/2}$.
\item \underline{Superhorizon limit $|k\tau|\ll 1$}. The mode equation now becomes
\begin{equation}
\frac{\d^2f}{\d y^2} + \frac{2}{y}\frac{\d f}{\d y} - \frac{2\kappa}{y^2}f = 0,
\end{equation}
where $y\equiv |k\tau|$. This can be solved analytically and we can express the general solution as
\begin{equation}
f = C_1\, y^{\left( -1 + \sqrt{1+8\kappa} \right) /2} + C_2\, y^{\left( -1 - \sqrt{1+8\kappa} \right) /2}.
\end{equation}
The second exponent is always smaller that the first one, so the second term dominates at late enough times (for $y \ll 1$). Recovering the physical field and scale factors, we have 
\begin{equation} \label{eq:parallel_perturbations_late_time_evolution}
{\delta\chi}^\parallel \simeq C^\prime_2\, 2^{1-\half\beta_-} \frac{H_I}{k^{3/2}} \left( \frac{k}{aH_I} \right)^{\half\beta_-}.
\end{equation}
The chosen normalization is motivated by the more detailed calculation\footnote{Note that here we use the redefined field $f = a {\delta X}^\parallel$ to keep the notation as slim as possible, whereas in App.~\ref{app:longitudinal_fluctuations}, where we present the full detailed calculation, we work with the field $\delta X^\parallel$, making it easier to determine the most useful normalization.} presented in App.~\ref{app:longitudinal_fluctuations}. 
Importantly, this superhorizon solution has the same time dependence as the solution for the homogeneous field and the transverse polarization.
\item \underline{Intermediate regime $|k\tau|\sim 1$}. Here we have no choice but to solve the full mode equation. As an analytic solution for it is hard to come by in this regime, we solve it numerically to connect the sub- and superhorizon expressions and determine $C_2^\prime$. As expected, we find a dependence of $C_2^\prime$ on $\kappa$ and we use the numerical solution to determine a fitting function $C_2^\prime \equiv |f(\kappa)|$. A particular example for the numerical evaluation and the matching of the analytic asymptotic results in the case of $\kappa = 1$ is shown in the left panel of Fig.~\ref{fig:numerical_interpolation} (cf. App.~\ref{app:longitudinal_fluctuations} for the relevant formulas).
Extracting several values of $C_2^\prime$ in the range $\kappa \in [10^{-4}, 10]$, which extensively covers our region of interest, we determine an accurate fit to the numerical results. It is given by
\begin{equation}\label{eq:kappa_fit}
|f(\kappa)| = 0.502 \kappa^{-0.5} - 0.224 + 0.262 \kappa - 0.0411 \kappa^2 + 0.00654 \kappa^3,
\end{equation}
which is depicted together with the residuals in the right panel of Fig.~\ref{fig:kappa_fit}.
\begin{figure}[t!]
    \centering
    \begin{minipage}[t]{0.47\textwidth}
        \centering
        \includegraphics[width=1.0\textwidth]{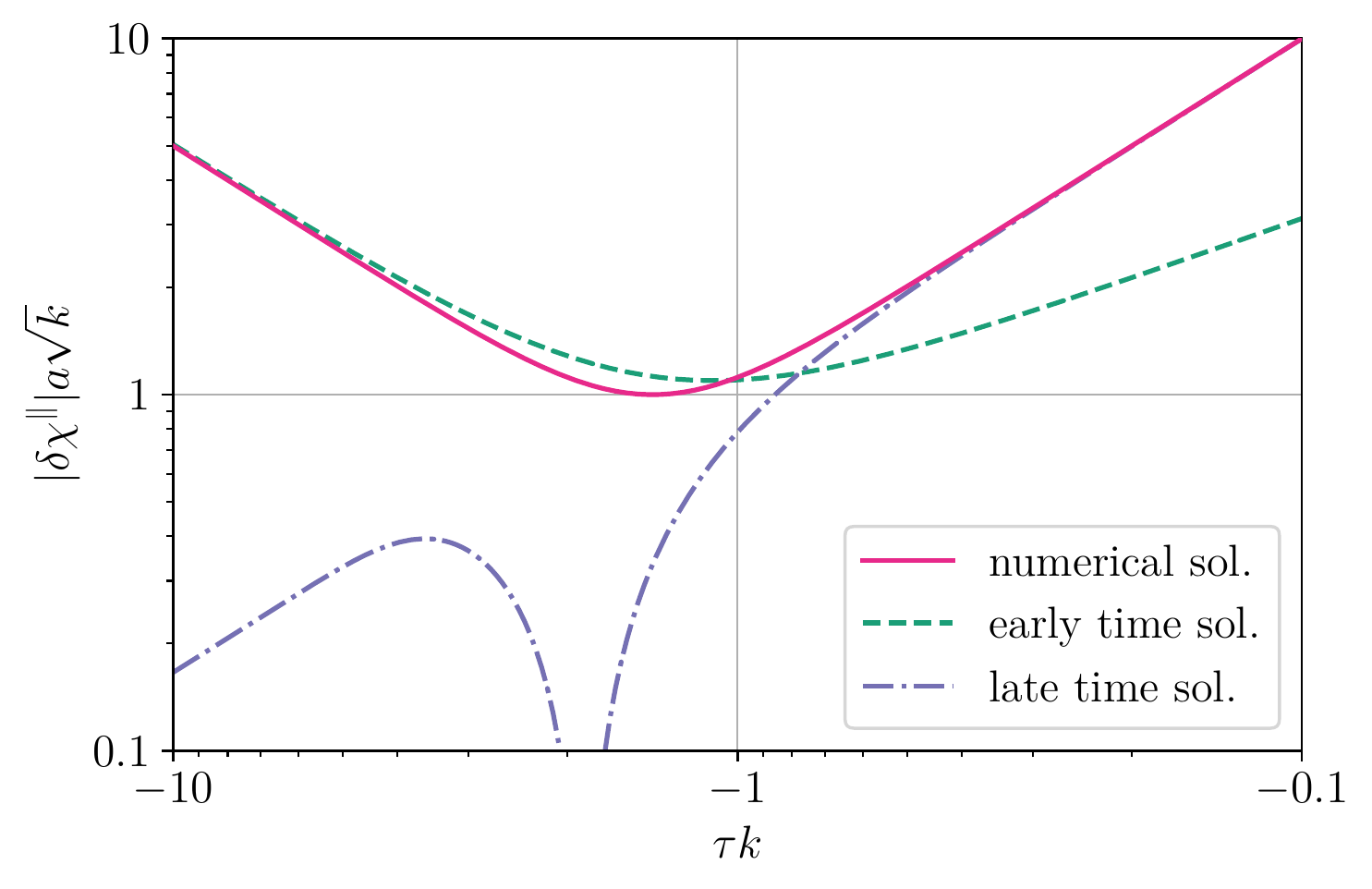}
        \caption{\footnotesize{Evolution of the early time, late time and numerical solution to the full mode equation in conformal time (cf. App.~\ref{app:longitudinal_fluctuations}) for $\kappa = 1$. The value of $C_2^\prime$ is chosen such that for late times the numerical solution agrees with the late time solution.}}
        \label{fig:numerical_interpolation}
    \end{minipage}\hfill
    \begin{minipage}[t]{0.47\textwidth}
        \hspace*{-0.5cm}
        \centering
        \includegraphics[width=1.0\textwidth]{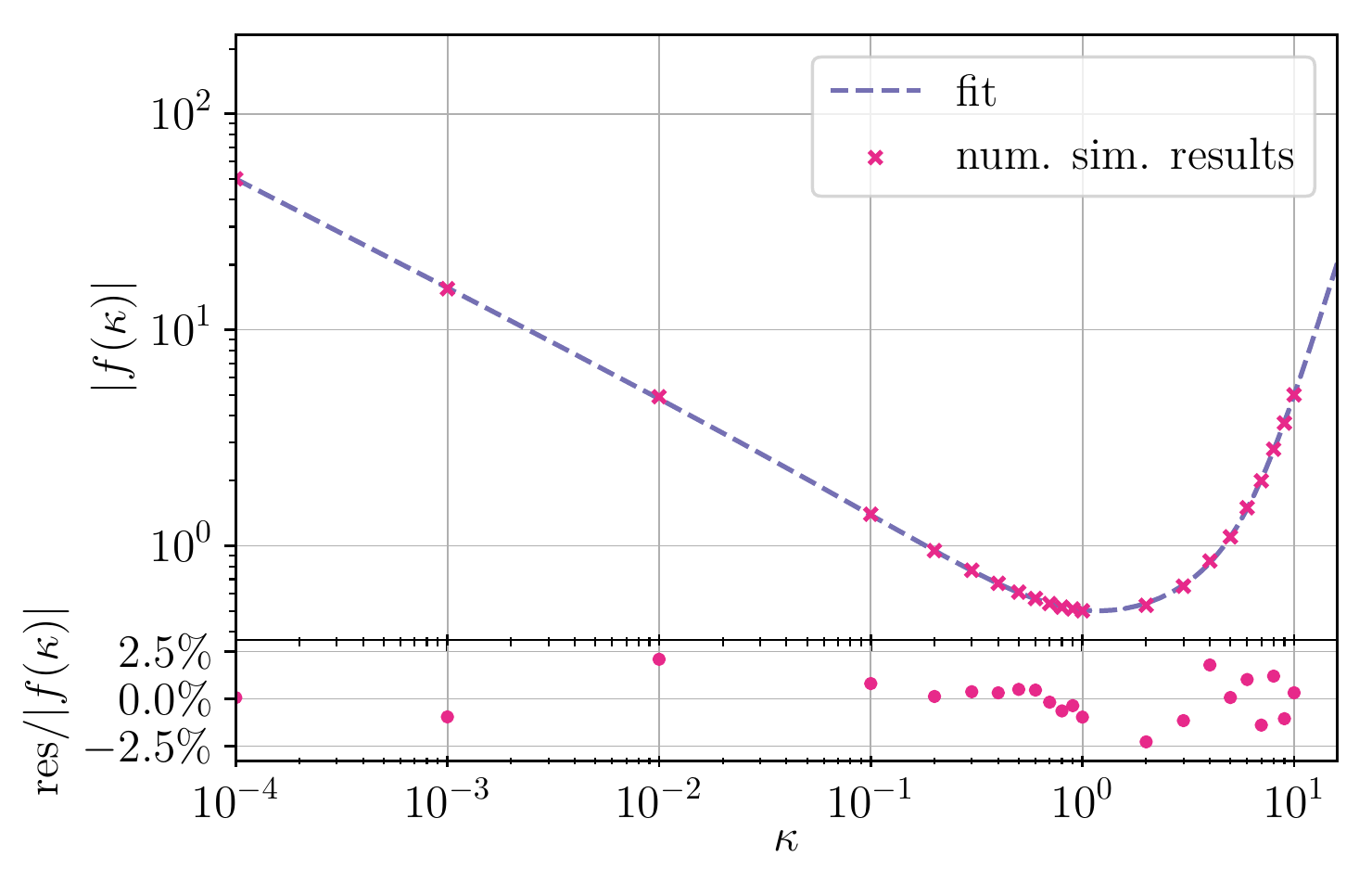}
        \caption{\footnotesize{Numerical results ($C_2^\prime$ values), determined from figures similar to Fig.~\ref{fig:numerical_interpolation}, and corresponding fit function $|f(\kappa)|$ shown together with the relative deviation of the fit function (residuals) from the numerical results underneath. The fit function for $|f(\kappa)|$ is given in Eq.~\eqref{eq:kappa_fit}.}}
        \label{fig:kappa_fit}
    \end{minipage}
\end{figure}
\end{itemize}
Putting it all together, the expression for the power spectrum of the longitudinal fluctuations can be written as
\begin{equation}
\mathcal{P}^\parallel_{\chi} (k,a) \simeq 2^{3-\beta_-} |f(\kappa)|^2 \left( \frac{H_I}{2 \pi} \right)^2 \left( \frac{k}{a H_I} \right)^{\beta_-},
\end{equation}
which is valid for superhorizon modes with $k \neq 0$. Comparing with the expression for the transverse fluctuations Eq.~\eqref{eq:ps_transverse_fluctiations_inflation}, we find the relation
\begin{equation}
\mathcal{P}^\parallel_{\chi} = \frac{2 \pi |f(\kappa)|^2}{\Gamma^2\!\left(\frac{3-\beta_-}{2}\right)}\, \mathcal{P}^\perp_{\chi_i}.
\end{equation}
This generalizes the result found in~\cite{Dimopoulos:2008yv,Karciauskas:2010as} for $\kappa = 1$, for which the relation $\mathcal{P}^\parallel_{\chi} \simeq 2\, \mathcal{P}^\perp_{\chi_i}$ is recovered.

We conclude that the primordial power spectrum of the longitudinal fluctuations is proportional to the one of the transverse fluctuations. In particular, for values of $\kappa$ close to unity, the proportionality factor is of $\mathcal{O}(1)$. However, as we will now see, the post-inflationary evolution of the two modes can significantly differ.


\subsubsection{Evolution after inflation}

We now discuss the evolution of the primordial fluctuation power spectrum generated during inflation throughout the different epochs in the history of the Universe until today. As in the previous section, we differentiate between transverse and longitudinal modes.

\medskip
\paragraph{Transverse fluctuations.}
The EOMs for the two perpendicularly polarized modes are
\begin{equation}\label{eq:EOM_transverse_perturbations}
\ddot{\delta\chi_i}^\perp + 3 H \dot{\delta\chi_i}^\perp + \left( m_X^2 + \frac{1 - \kappa}{6} R + \frac{k^2}{a^2} \right) {\delta\chi_i}^\perp = 0,
\end{equation}
which after the substitution $\kappa \rightarrow 1 - 6 \xi$ are the same as for the scalar perturbations: the transverse fluctuations thus evolve in the same way as the fluctuations of a scalar field. We conclude that, for large scales like the CMB ones, the ratio between the transverse fluctuations and the homogeneous field stays constant throughout the cosmological evolution. This implies that the density contrast power spectrum can be evaluated at any point, such as right after horizon exit. Doing so and adding up the two transverse polarizations, we find
\begin{equation}\label{eq:CMB_vector_power_spectrum}
\mathcal{P}^\perp_{\delta_{\chi}}(k_{\rm CMB}) \simeq \frac{8}{\chi_{e}^2\,\me^{\beta_-N_{\rm CMB}}} \left( \frac{H_I}{2\pi} \right)^2\, F(\beta_-).
\end{equation}
As expected, we encounter the same suppression (for $\kappa<1$) or enhancement (for $\kappa>1$) in the isocurvature perturbations as in the scalar case.

\medskip
\paragraph{Longitudinal fluctuations.}
The full EOM in this case is more complicated, but it greatly simplifies during the radiation domination epoch when $R=0$, or whenever we can neglect $R,\, \dot{R}\ll m_X^2$, yielding
\begin{equation}\label{eq:EOM_longitudinal_perturbations}
\ddot{\delta\chi}^\parallel +\left( 3 + \frac{2k^2}{k^2 + a^2m_X^2}\right) H \dot{\delta\chi}^\parallel + \left(\frac{2k^2}{k^2 +a^2m_X^2}H^2 + \frac{k^2}{a^2} + m_X^2 \right) {\delta\chi}^\parallel = 0.
\end{equation}
As long as reheating proceeds almost instantaneously, which we assume to be the case, this simplified version of the EOM is adequate to study the full post-inflationary evolution. Comparing Eq.~\eqref{eq:EOM_longitudinal_perturbations} with Eq.~\eqref{eq:EOM_transverse_perturbations}, it becomes clear that the behaviour of longitudinal perturbations can significantly differ from the one of transverse (and scalar) ones. Let us focus on large scale modes that become non-relativistic before they reenter the horizon, as these are the ones relevant for CMB observables. These modes evolve through three distinct regimes:

\begin{enumerate}[(i)]
\item $H \gg k/a \gg m_X$. In the superhorizon and relativistic limit, Eq.~\eqref{eq:EOM_longitudinal_perturbations} further simplifies to
\begin{equation}
\ddot{\delta\chi}^\parallel  + 5 H \dot{\delta\chi}^\parallel + 2H^2{\delta\chi}^\parallel = 0,
\end{equation}
which is solved by
\begin{equation}
{\delta\chi}^\parallel \simeq c_1^\prime a^{-1} + c_2^\prime a^{-2}.
\end{equation}
The physical field thus redshifts as $1/a$. The modes corresponding to the largest scales might not enter this regime at all if they become non-relativistic before the end of inflation, i.e. if they violate $k/a_r \gg m_X$. Therefore, the condition for the CMB modes to skip this regime is
\begin{equation}
m_X \gtrsim \frac{k_{\mathrm{CMB}}}{a_r} \simeq 2\cdot 10^{-4}\,\mathrm{eV}\,\sqrt{\frac{H_I}{6.6\cdot 10^{13}\,\mathrm{GeV}}},
\end{equation}
where $k_{\mathrm{CMB}}$ is a typical CMB scale, which we take to be the Plank pivot scale $k_{\rm CMB} = 0.05\,\mathrm{Mpc}^{-1}$, and $a_r$ is the scale factor at reheating, given by $a_r = \mathrm{e}^{-N_{\rm min}(H_I)}$ under the assumption of instantaneous reheating. We conclude that if $m\gtrsim 10^{-4}\,$eV, the CMB modes do not enter regime (i) if inflation happens at the highest scale compatible with the limit from the non observation of tensor modes~\cite{Akrami:2018odb}, $H_I<6.6\cdot 10^{13}\GeV$. For lower values of $H_I$, this condition shifts to lower and lower masses.

\item $H \gg m_X \gg k/a$.
Modes are superhorizon but already non-relativistic, which allows to approximate Eq.~\eqref{eq:EOM_longitudinal_perturbations} by
\begin{equation}
\ddot{\delta\chi}^\parallel  + 3 H \dot{\delta\chi}^\parallel = 0,
\end{equation}
and to obtain the solution
\begin{equation}
{\delta\chi}^\parallel \simeq c^{\prime\prime}_1 + c_2^{\prime\prime} a^{-1}.
\end{equation}
In principle, it seems that the constant mode would dominate the solution\footnote{Note that in terms of $\delta X_i^\parallel$ this would correspond to a growing mode. However, as the physical field is the one that controls the density perturbations, there is no real growth of fluctuations in this regime.}. In the absence of a non-minimal coupling, the authors of~\cite{Graham:2015rva} argue that this is actually not the case, because $c_1^{\prime\prime}\ll c_2^{\prime\prime}$ and the constant mode can be dropped. The reason for this is that in region (i) the solution becomes $\propto a^{-1}$ to great accuracy, and continuity of the solution and the first derivative imposes a large hierarchy between the two coefficients. This is not necessarily the case when a non-minimal coupling is present, because during inflation the superhorizon modes are not frozen and $\dot{\chi}_e\neq 0$ as Eq.~\eqref{eq:phi_e_dot} indicates. As a consequence, the ``flattening'' of the mode functions is either weakened or not applicable at all depending on how long the CMB modes spend in region (i), if they enter it at all.

On the one hand, if region (i) is skipped, that is, if the CMB modes become non-relativistic before the end of inflation, then there is no suppression. On the other hand, if $m_X \lesssim k_{\rm CMB} / a_r$ and the CMB modes enter region (i), i.e. they are still relativistic once radiation domination starts, then we expect some suppression of the power spectrum at those scales. By carefully matching the field value and its first derivative through the regimes (i) and (ii), we obtain that
\begin{equation}\label{eq:parallel_perturbations_suppression}
{\delta\chi}^\parallel(k_{\rm CMB},a_{\rm(ii)}) \simeq \left( \frac{m_X}{k_{\rm CMB}/a_r} \right)^{2} {\delta\chi}^\parallel(k_{\rm CMB},a_r),
\end{equation}
where $a_{\rm(ii)}$ denotes the scale factor at the end of region (ii).

\item $m_X \gg H,\, k/a$. Modes become massive enough to overcome Hubble friction, and we recover the usual pressureless-matter like EOM
\begin{equation}
  \ddot{\delta\chi}^\parallel + 3 H \dot{\delta\chi}^\parallel + m_X^2 {\delta\chi}^\parallel = 0,
\end{equation}
which is identical to the scalar case, as was already discussed for the homogeneous field and the transverse polarizations.
\end{enumerate}

\begin{figure}[t!]
\centering
\includegraphics[width=0.6\linewidth]{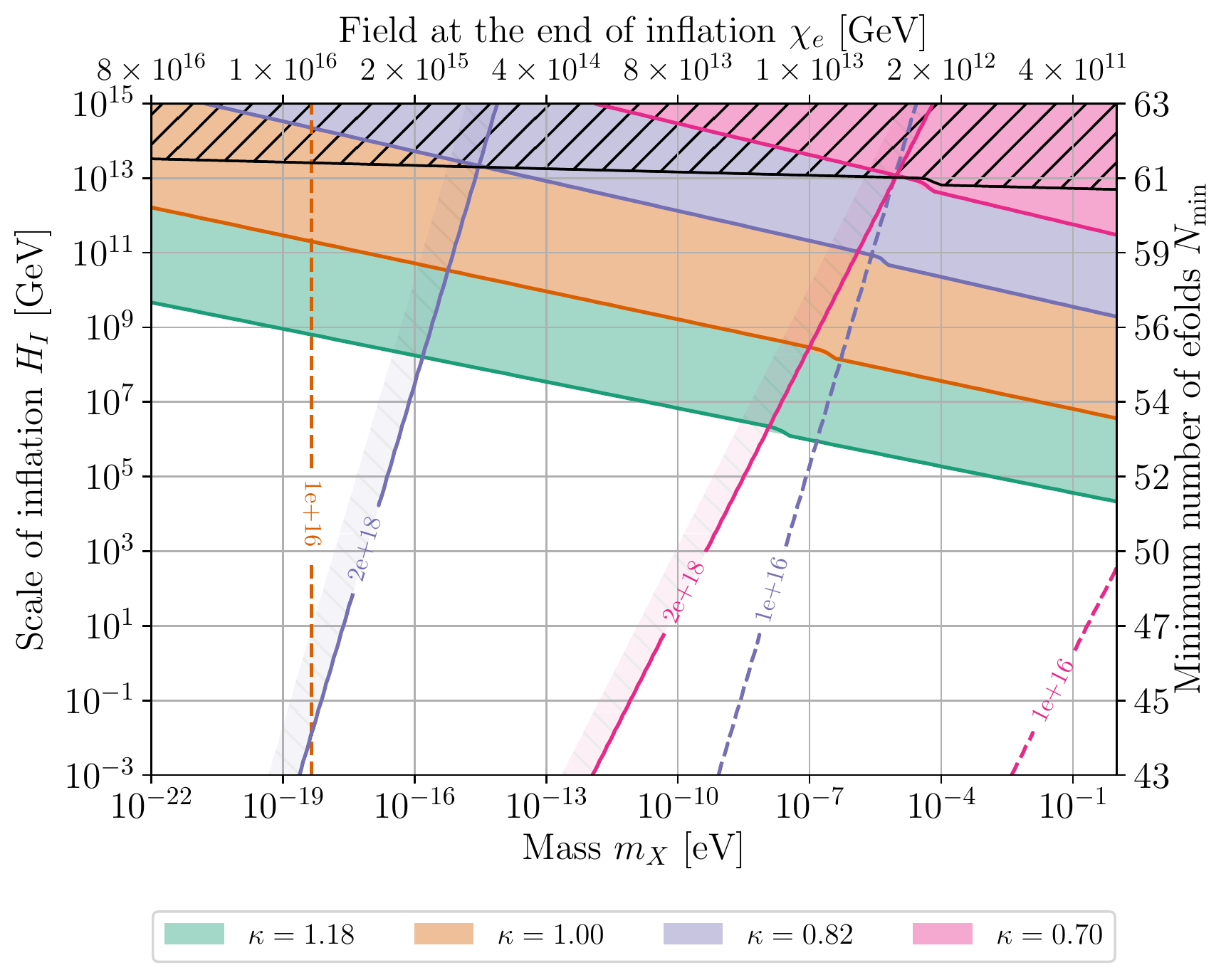}
\caption{\footnotesize{Parameter space showing the ability of a non-minimally coupled vector field to generate the observed dark matter abundance through the misalignment mechanism, as a function of its mass and the Hubble scale of inflation. This figure should be interpreted in the same way as Fig.~\ref{fig:isocurvature_scalars}. The kink in the isocurvature limits is due to the fact that longitudinal modes are only relevant for large enough masses, see Eq.~\eqref{eq:parallel_density_power_spectrum}.}}
\label{fig:isocurvature_vectors}
\end{figure}

We conclude that except for the potential suppression factor of Eq.~\eqref{eq:parallel_perturbations_suppression}, the evolution of longitudinal modes is analogous to that of the transverse ones, allowing us to write 
\begin{equation}\label{eq:parallel_density_power_spectrum}
\mathcal{P}^\parallel_{\delta_\chi}(k_{\rm CMB}) \simeq \frac{1}{2}\,\mathcal{P}^\perp_{\delta_\chi}(k_{\rm CMB}) \cdot \frac{2\pi |f(\kappa)|^2}{\Gamma^2\!\left(\frac{3-\beta_-}{2}\right)} \cdot \begin{cases}
1, &\mathrm{if}\ m\gtrsim k_{\rm CMB}/a_r\\
\left( \frac{m}{k_{\rm CMB}/a_r} \right)^{4}, &\mathrm{if}\ m\lesssim k_{\rm CMB}/a_r.
\end{cases}
\end{equation}
The total density contrast is the sum of the contributions from all the polarizations, so that
\begin{equation}\label{eq:vector_total_power_spectrum}
\mathcal{P}_{\delta_\chi}(k_{\rm CMB}) = \mathcal{P}^\perp_{\delta_\chi}(k_{\rm CMB}) + \mathcal{P}^\parallel_{\delta_\chi}(k_{\rm CMB}).
\end{equation}
In analogy to the scalar case, Fig.~\ref{fig:isocurvature_vectors} shows the Planck~\cite{Akrami:2018odb} isocurvature constraints in the $(m_X, H_I)$ parameter space for selected values of $\kappa$. The conclusion is similar to the scalar case: as long as $H_I\lesssim 10^{13}\GeV$, vector dark matter produced from the misalignament mechanism is compatible with isocurvature constraints provided a non-minimal coupling $\kappa\lesssim 1$ is present. 


\subsection{Energy density in fluctuations}
\label{sec:fluctuations_vectors}
Given the correspondence between transverse vector fluctuations and scalar ones, it is expected that the energy density stored in small scale fluctuations can also be important here. In addition to that and as a particularity of the vector case, the authors of~\cite{Graham:2015rva} (see~\cite{Ema:2019yrd} as well) showed that longitudinal vector fluctuations can be copiously produced during inflation also for $\kappa=0$.

\begin{figure}[t!]
\centering
\includegraphics[width=0.6\linewidth]{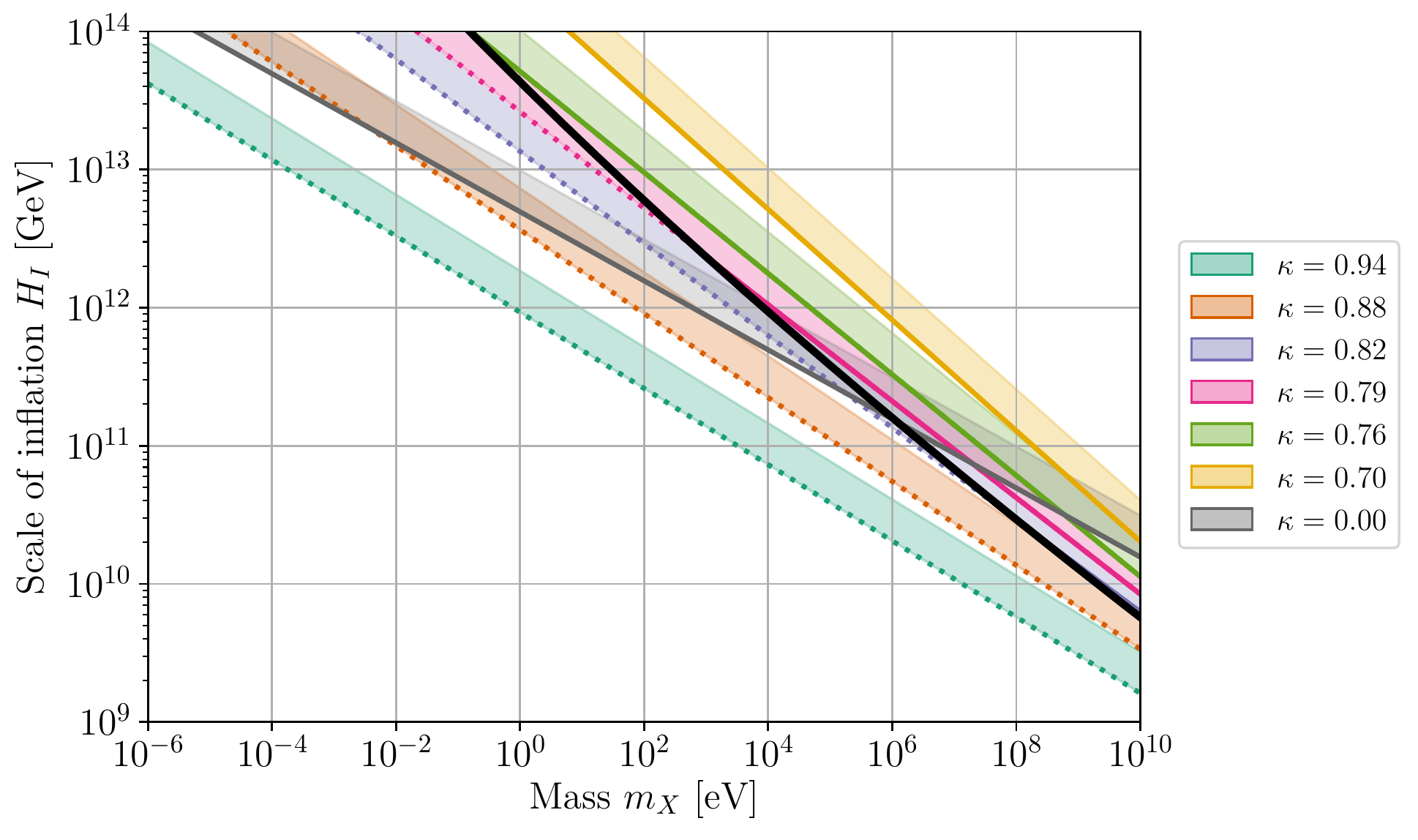}
\caption{\footnotesize{Regions of parameter space where the energy density stored in vector field fluctuations of inflationary origin makes a sizeable contribution to dark matter. This figure should be interpreted in the same way as Fig.~\ref{fig:fluctuations_density_isocurvature_scalars}, with the addition of the grey line, which corresponds to the contribution of longitudinal fluctuations in the case of a minimal coupling $\kappa=0$, as was computed in~\cite{Graham:2015rva} (the presence of a small enough non-minimal coupling as in Eq.~\eqref{eq:longitudinal_isocurvature_limit} is permitted; in this case the isocurvature constraints are also different and the full grey line is viable). We remind the reader that this figure assumes no contribution to the relic density from the misalignment mechanism.}}
\label{fig:fluctuations_density_isocurvature_vectors}
\end{figure}

As already discussed, the behaviour of the transverse modes is exactly the same as that of a scalar field after the identification $(1-\kappa)/6\leftrightarrow \xi$. At the same time, when $\kappa\sim 1$, the contribution from longitudinal modes at small scales around the peak at $k_{\star}^{-1}$ is suppressed due to the particularities of the post inflationary evolution described in the previous section. The easiest way to understand why the suppression at small scales is even stronger than at CMB scales is by replacing $k_{\rm CMB}$ by the much larger $k_\star$ in Eq.~\eqref{eq:parallel_density_power_spectrum}. All this means that the discussion in Sec.~\ref{sec:isocurvature_vector} carries over: the inflationary (transverse) fluctuations of a non-minimally coupled vector field can carry enough energy density to reproduce the observed dark matter abundance. As can be seen in Fig.~\ref{fig:fluctuations_density_isocurvature_vectors}, the region of parameter space where this production is viable corresponds to relatively high-scale inflation, masses above roughly eV and non-minimal couplings $\kappa$ moderately smaller than 1.

The contribution from longitudinal modes is however relevant for minimally coupled fields, \textit{i.e.} when $\kappa=0$, as was discussed in~\cite{Graham:2015rva,Ema:2019yrd}. The relic density that is generated in this case is given by
\begin{equation}
    \frac{\Omega_{X}}{\Omega_{\rm DM}}\simeq \sqrt{\frac{m_X}{3\cdot 10^{-5}\eV}}\left( \frac{H_I}{6.6\cdot 10^{13}\GeV} \right)^2.
\end{equation}
This is in contrast to misalignment and transverse fluctuations, whose contribution is completely negligible in the minimally coupled case. 
Longitudinal modes are very strongly suppressed at CMB scales due to the smoothing of the field that occurs during inflation when $\kappa=0$. However, the presence of even a small non-minimal coupling can significantly weaken this suppression. This is due to the fact that the modes describing the fluctuations have an $\mathcal{O}(\kappa)$ time derivative at the end of inflation, similarly to the homogeneous field value (cf. Eq.~\eqref{eq:phi_e_dot}). Taking this into account and carefully matching the solutions of the EOMs in the different regimes, we find that the amplitude of the power spectrum at CMB scales can be approximately computed as
\begin{align}\label{eq:longitudinal_isocurvature_limit}
    \frac{\mathcal{P}_\delta(k_{\rm CMB})}{\mathcal{P}_\delta(k_\star)} \simeq \kappa^2 \left( \frac{k_{\rm CMB}}{m_X \cdot a_r} \right)^{2-4\kappa}
    \simeq 1.7\cdot 10^{3}\,\kappa^2 \left( \frac{6\cdot 10^{-6}\eV}{m_X} \right)^{\frac{9}{4}(1-2\kappa)},
\end{align}
which is valid for sufficiently small but non-vanishing $\kappa$. Taking into account that the power spectrum is of $\mathcal{O}(1)$ at the peak, \textit{i.e.} $\mathcal{P}_\delta(k_\star)\sim 1$, and applying the Planck constraint~\cite{Akrami:2018odb} on isocurvature perturbations $\mathcal{P}_\delta(k_{\rm CMB}) < 1.3\cdot 10^{-9}$, we conclude that the non-minimal coupling is bound to be very small if the vector is light. For instance, a mass of $m_X \sim 10^{-6}\eV$ requires $\kappa\lesssim 10^{-6}$. If this condition is satisfied, a high inflationary scale allows for generation of sub-eV vector dark matter in this case, as can be seen in Fig.~\ref{fig:fluctuations_density_isocurvature_vectors}.


\subsection{Stochastic scenario}
\label{sec:stochastic_scenario_vector}

In analogy to the scalar case discussed in Sec.~\ref{sec:stochastic_scenario_scalar}, the stochastic scenario can also be realized for a vector field with a non-minimal coupling to gravity. Once more, the calculation is simplified by considering the contribution from transverse and longitudinal fluctuations separately. As expected, transverse modes behave like scalar ones with the usual replacements $\alpha_\pm \leftrightarrow \beta_\pm$ and $\xi \leftrightarrow (1 - \kappa)/6$. Following App.~\ref{app:stochastic_scenario} and once equilibrium is reached, the variance of the transverse superhorizon fluctuations is given by
\begin{equation}\label{eq:stochastic_transverse_variance}
\left\langle {{\chi}^\perp_\delta}^2 \right\rangle \simeq 2\, \frac{F(\beta_-)}{\beta_-} \left(\frac{H_I}{2 \pi}\right)^2
\mathrm{e}^{-\beta_- N_\mathrm{min}},
\end{equation}
where the factor of 2 accounts for the two transversal polarizations.
For the longitudinal fluctuations, we make use of the results of Sec.~\ref{sec:isocurvature_vector} to write the mode functions as
\begin{equation}
{\delta\chi}^\parallel \simeq |f(\kappa)|\, 2^{1-\half\beta_-}\, \frac{H_I}{k^{3/2}} \left( \frac{k}{aH_I} \right)^{\half\beta_-},
\end{equation}
which is valid after horizon exit during inflation. The variance of the Gaussian distribution of the longitudinal field fluctuations can also be found by integrating over the superhorizon contributions\footnote{Note that the integral Eq.~\eqref{eq:stochastic_integral} is dominated by the upper limit, which means that the variance is dominated by the smallest superhorizon modes.}.
At the end of inflation, both transverse and longitudinal contributions are of the same size. However, the post-inflationary evolution is different for each polarization. Because of the flattening of the wavefunctions described in Sec.~\ref{sec:isocurvature_vector} (cf. also~\cite{Graham:2015rva}), longitudinal modes will be suppressed unless
\begin{equation}
    m_X\gtrsim \frac{k_0}{a_r} \sim 10^{-7}\eV\,\sqrt{\frac{6.6\cdot 10^{13}\GeV}{H_I}},
\end{equation}
where $k_0^{-1}\sim 10\,\mathrm{Gpc}$ is the size of the observable Universe. The variance of superhorizon longitudinal fluctuations is thus
\begin{equation}\label{eq:stochastic_longitudinal_variance}
\left\langle {{\chi}^\parallel_\delta}^2 \right\rangle \simeq \frac{1}{2}\,\left\langle {{\chi}^\perp_\delta}^2 \right\rangle \cdot \frac{2\pi |f(\kappa)|^2}{\Gamma^2\!\left(\frac{3-\beta_-}{2}\right)} \cdot \begin{cases}
1, &\mathrm{if}\ m\gtrsim k_{0}/a_r\\
\left( \frac{m}{k_{0}/a_r} \right)^{4}, &\mathrm{if}\ m\lesssim k_{0}/a_r.
\end{cases}
\end{equation}
The total variance is obtained by adding up the contributions from both polarizations,
\begin{equation}\left\langle {{\chi}_\delta}^2 \right\rangle = \left\langle {{\chi}^\perp_\delta}^2 \right\rangle + \left\langle {{\chi}^\parallel_\delta}^2 \right\rangle,
\end{equation}
so that the typical field value is given in the stochastic regime by $\sqrt{\left\langle {{\chi}_\delta}^2 \right\rangle}$.

As in the scalar case, the stochastic regime is only reached if inflation lasts for an extra $\Delta N$ e-folds in addition to $N_{\rm min}$. On the one hand, $\Delta N > 1/\beta_-$ is required for the random walk process to attain equilibrium. On the other hand, dominance (by a factor of $\gamma$) of the variance of the stochastic distribution over any remnant of the initial field value $\chi_s$ sets the additional constraint
\begin{equation}
\Delta N > \frac{1}{\beta_-} \ln\left[ \gamma \left( \frac{\pi |f(\kappa)|^2}{\Gamma^2\!\left( \frac{3 - \beta_-}{2} \right)} + 1 \right)^{-1} \frac{2^{1 + \beta_-} \pi^3}{\Gamma^2\!\left( \frac{3 - \beta_-}{2} \right)} \frac{\beta_-}{(\beta_+ - \beta_-)^2} \left( \frac{\dot{\phi_s}}{H_I^2} + \frac{1}{2} \beta_+ \frac{\phi_s}{H_I} \right)^2 + 1 \right].
\end{equation}
Because the stochastic scenario does not modify the last $N_{\rm min}$ e-folds of inflation, isocurvature constraints can be computed in the same way as for the misalignment mechanism. Using Eqs.~\eqref{eq:vector_total_power_spectrum},~\eqref{eq:stochastic_transverse_variance} and~\eqref{eq:stochastic_longitudinal_variance}, we find that the density contrast power spectrum at CMB scales is given by
\begin{equation}
    \mathcal{P}_\delta(k_{\rm CMB}) \sim 4\,\beta_-\,\me^{\beta_-(N_{\rm min} - N_{\rm CMB})},
\end{equation}
where we have neglected a small effect coming from the different scale dependence ($k_{0}$ vs $k_{\rm CMB}$) of the longitudinal power spectrum: see the difference between Eq.~\eqref{eq:stochastic_longitudinal_variance} and Eq.~\eqref{eq:parallel_density_power_spectrum}. This equation highlights the fact that the accumulated variance is enhanced by a factor of $1/\beta_-$ with respect to that of each individual mode. The need for a strong suppression of the CMB isocurvature makes it hard to realize a stochastic scenario for $\kappa\neq 1$. Indeed, the Planck~\cite{Akrami:2018odb} constraint translates into a limit $1-\kappa\lesssim 10^{-10}$ on the deviation of the non-minimal coupling from the scalar-like value. 


\section{Conclusions and outlook}
\label{sec:conclusions}


In the present paper we have considered light (pseudo-)scalar and vector fields with couplings to the curvature/Ricci scalar as candidates for dark matter. 
The main impact of these couplings occurs during inflation when they give a contribution to the mass of the fields. The effective mass is then typically of the order of the Hubble scale, leading to a non-trivial evolution of both the field and the fluctuations during inflation. 
This evolution during inflation impacts all three possible non-thermal scenarios for the production of the dark matter density: misalignment~\cite{Preskill:1982cy,Abbott:1982af,Dine:1982ah,Nelson:2011sf,Arias:2012az,Jaeckel:2013uva}, stochastic~\cite{Graham:2018jyp,Guth:2018hsa,Ho:2019ayl,Tenkanen:2019aij} and fluctuation~\cite{Graham:2015rva,Nurmi:2015ema,Kainulainen:2016vzv,Bertolami:2016ywc,Cosme:2018nly,Alonso-Alvarez:2018tus} production.
After inflation, the Ricci scalar vanishes during radiation domination and the evolution proceeds in the standard way.

For (pseudo-)scalars, even relatively small positive vales of $\xi\lesssim {\mathcal{O}}(0.1)$ open up sizeable areas in parameter space for the \emph{misalignement} mechanism due to the suppression of isocurvature fluctuations at the CMB scales.  
In the opposite direction, negative values of $\xi\lesssim -4$ are excluded or require stabilization by an additional term in the potential because the induced tachyonic mass produces trans-Planckian field excursions, even for the smallest possible values of the inflationary Hubble scale (for high-scale inflation, this occurs for much smaller couplings, $\xi\lesssim -0.1$). 

For high scale inflation and larger masses above $\sim 1\eV$, the energy density stored in small-scale fluctuations of the scalar field becomes sizeable up to the point where it can account for the entire observed dark matter density. In this situation, dark matter is produced from \emph{fluctuations} of the field generated during inflation, long after the CMB modes exit the horizon. The presence of a non-minimal coupling $\xi \gtrsim 0.03$ suppresses fluctuations at large scales, avoiding the Planck isocurvature constraints and leading to a peaked density contrast power spectrum~\cite{Alonso-Alvarez:2018tus}. 

Finally, assuming that inflation lasts for a sufficiently long time, the accumulation of superhorizon fluctuations with wavelengths larger than the size of the visible Universe can dominate the observable homogeneous field over any initial condition. In this stochastic scenario, the homogeneous field value for the post-inflationary evolution in our Hubble patch is chosen probabilistically from a Gaussian distribution whose variance is controlled by the curvature coupling $\xi$. In order for the super Hubble variance to grow sufficiently large and avoid isocurvature constraints, the minimal coupling is required to be very small, $\xi\lesssim 10^{-10}$, while the number of e-folds of inflation has to be larger than $\sim 1/(8\xi)\sim 10^{9}$.

A massive vector can also act as dark matter and is subject to production through the same three mechanisms as a scalar. The cosmological evolution of the homogeneous component of a vector field with curvature coupling $\kappa$ is found to mimic that of the scalar field after the identification $\kappa \leftrightarrow 1-6\xi$. Consequently, the misalignment mechanism proceeds in the usual way, with the distinction that the minimally coupled scalar corresponds to a non-minimally coupled vector with $\kappa = 1$. It should however be noted that the massive vector acquires a longitudinal polarization whose kinetic term is rendered negative for a finite range of momenta due to the presence of the non-minimal coupling. A detailed study of how this sign flip affects the stability of the vacuum, along the lines of~\cite{Himmetoglu:2008zp,Himmetoglu:2009qi,Karciauskas:2010as,Nakayama:2019rhg}, is left for future work.

The discussion of vector fluctuations is most easily carried out by considering transverse and longitudinal polarizations separately. As for the homogeneous field and due to their approximate conformal symmetry, transverse fluctuations behave like scalar fluctuations with the identification of curvature couplings mentioned in the previous paragraph. This means that in the misaligned regime, isocurvature constraints are weakened\footnote{A tachyonic mass for the vector is generated during inflation if $\kappa>1$. In analogy to the scalar case, avoiding trans-Planckian field excursions sets strong constraints on the size of the curvature coupling.} if $\kappa\lesssim 1$. The phenomenologically interesting region where misalignment production is viable even for high scale inflation corresponds to $1-\kappa$ being $\mathcal{O}(0.1)$. The vector DM production from transverse fluctuations and the stochastic scenario are also viable and proceed in a manner analogous to the scalar case.

The longitudinal mode, however, presents a number of particularities with respect to the transverse ones. In the misalignment mechanism and owing to the differences both in its inflationary and post-inflationary evolution, the longitudinal mode is shown to be suppressed if the vector is very light. However, when it is heavier than $10^{-4}\eV$ (or less depending on the scale of inflation), the contribution of longitudinal modes to the isocurvature spectrum at CMB scales is of the same size as that of the transverse ones and cannot be neglected. The situation of the production from inflationary fluctuations is also interesting. While small-scale longitudinal fluctuations are always suppressed when $\kappa\sim 1$, in the case of a minimally coupled vector their amplitude can be large enough to contain an energy density comparable to that of the DM, provided inflation occurs at a high scale. We show that the viability of this scenario strictly relies on the almost complete absence of a non-minimal coupling; for instance, a $\kappa\gtrsim 10^{-6}$ spoils the mechanism if $m_X\sim 10^{-6}\eV$.

Throughout this paper we have considered the inflationary epoch of our Universe to be a perfect de Sitter expansion with constant $H_I$. However, the observation of a slight scale dependence in the fluctuation spectrum at CMB scales~\cite{Akrami:2018odb} indicates that $H_I$ is a slowly decreasing function of time. As long as this spectral tilt remains small\footnote{The spectral index of adiabatic modes has only been measured around the scales accessible in the CMB. Without more observational guidance, it is reasonable to make the assumption that $n_s$ remains small during most of inflation. A full discussion of the physical consequences of large variations in $H_I$ is an interesting topic, but one that lies beyond the scope of this work.}, the results of this work are not changed qualitatively. Quantitatively, the effect of the spectral tilt can be compensated by a shift to slightly larger (for scalars) or smaller (for vectors) values of the non-minimal coupling. 

Non minimal curvature couplings allow for the generation of light (pseudo-)scalar or vector dark matter in a wide range of masses. The three different production mechanisms
complement each other in terms of their viability in different regions of parameter space. This opens up exciting possibilities for experiments and potentially interesting interactions between the physics of dark matter and inflation.


\acknowledgments


We would like to thank Fuminobu Takahashi, Manuel Reichert, Tommi Markkanen and Tommi Tenkanen for interesting discussions. The work of G. A. is supported through a ``la Caixa” predoctoral grant of Fundaci\'on ``la Caixa''. This project has received funding from the European Union's Horizon 2020 research and innovation programme under the Marie Sklodowska-Curie grant agreement No 674896 (ITN ELUSIVES).


\appendix


\section{Inflation scale and minimal number of e-folds}
\label{app:length_of_inflation}
Being crucial to our calculations, here we give a somewhat exhaustive discussion about how long inflation needs to have lasted, based on observations, the standard cosmology and basic assumptions about the model of inflation and reheating. We will closely follow Ref.~\cite{Liddle:2003as} for this purpose.

Let us denote by $N(k)$ the number of e-folds at which the comoving scale $k$ left the horizon, that is, when $k=aH$ during inflation. Note that the requirement of an accelerated expansion implies that $aH$ is a monotonically increasing function of time. The largest scale to which we have access observationally is the present horizon scale $k_0 = a_0H_0$. The requirement $N(k_0)\leq N_{\mathrm{tot}}$ gives a lower limit on the total number of e-folds of inflation, but it cannot say anything about the actual value of $N_{\mathrm{tot}}$, which can (and usually is expected to) be much larger. In order to make numerical statements, we will often make the assumption of \emph{minimal inflation}, which sets $N_{\mathrm{tot}}$ to its smallest observationally allowed value $N_{\mathrm{min}}=N(k_0)$.

The general expression for $N(k)$ is~\cite{Liddle:2003as}
\begin{equation}\label{eq:number_efolds_k}
N(k) = -\ln\left( \frac{k}{a_0H_0} \right) + \frac{1}{3}\ln\left( \frac{\rho_{\mathrm{reh}}}{\rho_{\mathrm{end}}} \right) + \frac{1}{4}\ln\left( \frac{\rho_{\mathrm{eq}}}{\rho_{\mathrm{reh}}} \right) + \ln\left( \frac{H_{k}}{H_{\mathrm{eq}}} \right) + \ln \left(219\,\Omega_0 h\right),
\end{equation}
where $\rho_{\mathrm{end}},\,\rho_{\mathrm{reh}},\,\rho_{\mathrm{eq}}$ are the energy density at the end of inflation, at reheating and at matter-radiation equality, respectively; and $H_k$ is the Hubble parameter at the time when the scale $k$ exits the horizon during inflation. We can simplify this expression by making two main assumptions:
\begin{itemize}
\item $H_k\sim H_{\mathrm{end}}\sim H_I$: there is no energy drop during inflation. This amounts to assuming exactly exponential expansion with a constant value of $H$.
\item $\rho_{\mathrm{reh}}\sim\rho_{\mathrm{end}}$: reheating happens instantaneously when inflation ends.
\end{itemize}
Under these assumptions and using $\rho = 3 m_{\mathrm{pl}}^2 H^2$ (with the reduced Planck mass), $H_{\mathrm{eq}} \approx 2 \cdot 10^{-27} \,\mathrm{GeV}$, the matter density $\Omega_0 \approx 0.3$ and $h \approx 0.7$, we can obtain a simple expression for the minimum number of e-folds, $N_{\mathrm{min}}$, which reads
\begin{equation}\label{eq:N_min}
N_{\mathrm{min}} = 61.97 + \half \ln\left( \frac{H_I}{6.6\cdot 10^{13}\,\mathrm{GeV}} \right).
\end{equation}
We normalize the result to the largest scale of inflation allowed by observations (using the $95\%$ cl limit from~\cite{Akrami:2018odb}). Note that $N_{\mathrm{min}}$ becomes smaller for lower inflationary scales. Another quantity that is relevant for us is $N(k_{\mathrm{CMB}})$, which is the number of e-folds at which the perturbations accesible in the CMB exited the horizon. Using the Planck pivot scale $k_{\mathrm{CMB}}=k_{\star} = 0.05\,\mathrm{Mpc}^{-1}$, we obtain that these scales exited the horizon
\begin{equation}
N_{\mathrm{min}} - N(k_{\mathrm{CMB}}) = 7.26
\end{equation}
e-folds after the current horizon scale. Finally, let us relax the two former assumptions and give a more general expression for the number of e-folds,
\begin{equation}
N_{\mathrm{min}} = 61.97 + \half \ln\left( \frac{H_I}{6.6\cdot 10^{13}\,\mathrm{GeV}} \right) + \frac{1}{4}\ln\left( \frac{V_{k_0}}{\rho_{\mathrm{end}}} \right) + \frac{1}{12}\ln\left( \frac{\rho_{\mathrm{reh}}}{\rho_{\mathrm{end}}} \right).
\end{equation}
Here, we use the slow-roll approximation to write $H_k = 8\pi V_k / 3\mpl^3$ and express the result in terms of the energy scale of inflation instead of the Hubble scale. Note that a longer period of preheating generally implies $\rho_{\mathrm{reh}}\leq\rho_{\mathrm{end}}$ and leads to a decrease in $N_{\mathrm{min}}$, while a deviation from pure de Sitter expansion means that $V_{k_0}\geq\rho_{\mathrm{end}}$ and results in a larger value for $N_{\mathrm{min}}$.


\section{Stochastic scenario}
\label{app:stochastic_scenario}

One aspect that is relevant to the stochastic scenario is the exact form of the field value evolution during the inflationary period. For arbitrary initial conditions $\phi_s$ and $\dot{\phi}_s$, it is given by
\begin{equation}
\phi(t) = \phi_s \left( c_1 \me^{-\half \alpha_- H_I t} + c_2 \me^{-\half \alpha_+ H_I t} \right),
\end{equation}
with 
\begin{equation}\label{eq:c1_c2}
\begin{aligned}
c_1 &= \frac{1}{\alpha_+ - \alpha_-} \left( \alpha_+ + \frac{\dot{\phi}_s}{H_I\,\phi_s} \right),\\
c_2 &= - \frac{1}{\alpha_+ - \alpha_-} \left( \alpha_- + \frac{\dot{\phi}_s}{H_I\,\phi_s} \right),
\end{aligned}
\end{equation}
and identical equations being valid in the vector case for $\chi_i(t)$ only exchanging $\alpha_\pm$ for $\beta_\pm$. Using that $\alpha_+ > \alpha_-$ ($\beta_+ > \beta_-$), we find for late times the expressions
\begin{align}\label{eq:app_scalar_field_evolution}
\phi(t) &\approx \frac{2}{H_I (\alpha_+ - \alpha_-)} \left( \dot{\phi}_s + \frac{1}{2} \alpha_+ H_I \phi_s \right) \mathrm{e}^{-\frac{1}{2} \alpha_- H_I t} \\
\dot{\phi}(t) &\approx -\frac{\alpha_-}{\alpha_+ - \alpha_-} \left( \dot{\phi}_s + \frac{1}{2} \alpha_+ H_I \phi_s \right) \mathrm{e}^{-\frac{1}{2} \alpha_- H_I t}.
\end{align}
Note that $t$ is defined such that at the end of inflation $H_I t = N_\mathrm{tot}$. Furthermore, we can see from these expressions that in general
\begin{equation}
\dot{\phi}(t) = -\frac{1}{2} \alpha_- H_I \phi(t),
\end{equation}
so in terms of initial conditions the assumption $\dot{\phi}_s \sim H_I \phi_s$ is reasonable as long as $\alpha_- \not\approx 0$.

Let us now turn into the stochastic regime, which involves the computation of the variance of the field induced by the non-vanishing momentum modes that describe the fluctuations. As noted in Sec.~\ref{sec:stochastic_scenario_scalar}, our calculation is done along the lines of~\cite{Guth:2018hsa}. There, the quantity $\nu$ appears, which is given by $\alpha_- = 3 - 2 \nu$ in terms of the quantities used throughout this paper, or more directly by $\nu \approx 3/2 \sqrt{1 - 16/3 \xi}$, with $0 \leq \xi \leq 3/16$.
As in~\cite{Guth:2018hsa}, we can also express the Hankel functions that describe the modes in terms of Bessel functions and see that the contribution of $J_{-\nu}(\cdot)$ is the dominant one, leading to
\begin{equation}
|\delta \phi_k|^2 \simeq \frac{H_I^2}{4 \pi}\, \Gamma^2\!\left(\frac{3-\alpha_{-}}{2}\right) \left( \frac{1}{a H_I} \right)^3 \left( \frac{2 a H_I}{k} \right)^{3-\alpha_{-}}.
\end{equation}
Here, the subscript ``horizon'' indicates the quantity at horizon exit.
These fluctuations contribute to our Universe's homogeneous field value as long as inflation stretches to scales that are still superhorizon today. The accumulated effect of all the sufficiently long wavelength modes results in a Gaussian distribution for the homogeneous field value, with variance
\begin{equation}\label{eq:stochastic_integral}
\begin{aligned}
\left\langle \phi_\delta^2 \right\rangle_{\rm horizon} &= \int_{a(t_i) H_I}^{a(t_\mathrm{horizon}) H_I} \frac{\d^3 \Vec{k}}{(2 \pi)^3}\, |\delta \phi_k|^2_{a = a_\mathrm{horizon}} \\
&= \frac{2^{-\alpha}}{\pi^3 \alpha_-} \Gamma^2\!\left( \frac{3 - \alpha_-}{2} \right) H_I^2 \left[ 1 - \mathrm{e}^{-\alpha_- (N_\mathrm{tot} - N_\mathrm{min})} \right].
\end{aligned}
\end{equation}
The field value right at horizon exit is randomly drawn from this Gaussian distribution. Assuming that it is centered at the origin, the ``typical'' result  corresponds to $\phi_\mathrm{horizon} = \sqrt{\left\langle \phi_\delta^2 \right\rangle}$. That said, this homogeneous field can vary significantly by the end of inflation due to the potentially fast superhorizon evolution caused by the non-minimal coupling. We can use Eq.~\eqref{eq:app_scalar_field_evolution} with $\dot{\phi}_\mathrm{horizon} = 0$ (this assumes that equilibrium has been reached in the sense that is described below) to describe the subsequent evolution after horizon exit by
\begin{equation}
\phi_{e} \approx \frac{\alpha_+}{\alpha_+ - \alpha_-}\, \mathrm{e}^{-\frac{1}{2} \alpha_- N_\mathrm{min}} \phi_\mathrm{horizon}.
\end{equation}
Taking into account this superhorizon evolution during the last $N_{\rm min}$ efolds of inflation, the typical homogeneous field value at the end of inflation in our Hubble patch is
\begin{equation}
\label{eq:stochasticfull}
\left\langle \phi_\delta^2 \right\rangle = \frac{F(\alpha_-)}{\alpha_-} \left(\frac{H_I}{2 \pi}\right)^2 \left[ 1 - \mathrm{e}^{-\alpha_- (N_\mathrm{tot} - N_\mathrm{min})} \right] \left( \frac{\alpha_+}{\alpha_+ - \alpha_-}\, \mathrm{e}^{-\frac{1}{2} \alpha_- N_\mathrm{min}} \right)^2.
\end{equation}
The truly stochastic regime, when the distribution is well approximated by a Gaussian, is only reached after a sufficient amount of inflation. This occurs when $\Delta N = N_\mathrm{tot} - N_\mathrm{min} > 1/\alpha_-$ so that the integral Eq.~\eqref{eq:stochastic_integral} is dominated by the superhorizon modes with the shortest wavelengths. In this regime, the term in the square brackets in Eq.~\eqref{eq:stochasticfull} can be dropped. By also using $\alpha_+ / (\alpha_+ - \alpha_-) \approx 1$ we get the result given in Sec.~\ref{sec:stochastic_scenario_scalar}.

The stochastic contribution to the homogeneous field competes with the exponentially decaying contribution from the initial field value at the start of inflation $\phi_s$. Determining the extra number of e-folds (before modes dominating Eq.~\eqref{eq:stochastic_integral} exit the horizon) that are necessary for the variance in Eq.~\eqref{eq:stochasticfull} to be a factor of $\gamma$ bigger than the contribution from initial condition, cf. Eq.~\eqref{eq:app_scalar_field_evolution}) is straightforward.  To do so we have to ensure that
the remnant of the initially homogeneous field value is a factor of $\gamma$ smaller than the variance. The result is given in Eq.~\eqref{eq:stochastic_scalar_N}.

The calculations for the stochastic scenario in the case of a vector field go along the same lines as the ones presented above, with the contributions being split into transverse and longitudinal modes.


\section{Longitudinal fluctuations}
\label{app:longitudinal_fluctuations}

The EOMs for the longitudinal modes of the physical field $\chi$ in momentum space (cf.~\cite{thomasMaster}) are
\begin{align} \label{eq:vector_longitudinal_perturbations_eom}
\begin{aligned}
0 &= \ddot{\delta\chi}^\parallel + \left[ 3 H + \frac{k^2}{k^2 + a^2 \left( m_X^2 - \frac{\kappa}{6} R \right)} \left( 2 H - \frac{\kappa}{6} \frac{\dot{R}}{m_X^2 - \frac{\kappa}{6} R} \right) \right] \dot{\delta\chi}^\parallel \\
&\quad + \left[ m_X^2 + \frac{1 - \kappa}{6} R + \frac{k^2}{a^2} + \frac{k^2 H}{k^2 + a^2 \left( m_X^2 - \frac{\kappa}{6} R \right)} \left( 2 H - \frac{\kappa}{6} \frac{\dot{R}}{m_X^2 - \frac{\kappa}{6} R} \right) \right] \delta\chi^\parallel.
\end{aligned}
\end{align}
When specifying to inflation, it turns out to be advantageous to work with the original field $\delta X^\parallel = a \, \delta \chi^\parallel$ and to switch to conformal time using $\mathrm{d}t / \mathrm{d}\tau = a$ and $a = - 1 / (\tau H_I)$. This deviates from the redefinition used in the main text, $f = a \delta X^\parallel$, but it allows to more easily obtain more precise analytic approximations of the mode functions. With this, we find (cf.~\cite{thomasMaster})
\begin{equation} \label{eq:vector_eom_inflation}
0 = \left[ \partial_\tau^2 - \frac{2 \tau k^2 H_I^2}{\tau^2 k^2 H_I^2 + m_X^2 - 2 \kappa H_I^2} \partial_\tau + \frac{m_X^2 - 2 \kappa H_I^2}{\tau^2 H_I^2} + k^2 \right] \delta X^\parallel.
\end{equation}
This equation can be simplified and analytically solved in the limits of interest to us.

In the subhorizon limit, when $k / (a H_I) = -k\tau \gg 1$, we can approximate Eq.~\eqref{eq:vector_eom_inflation} by
\begin{equation}
0 = \left[ \partial_\tau^2 - \frac{2}{\tau} \partial_\tau + \frac{\widetilde{m}^2}{\tau^2 H_I^2} + k^2 \right] \delta X^\parallel,
\end{equation}
where we introduced the shifted mass $\widetilde{m}^2 \equiv m_X^2 - 2 \kappa H_I^2$. As is done in~\cite{Dimopoulos:2008yv}, the additional redefinition $\widetilde{\delta X}_i^\parallel \equiv - |\widetilde{m}| / (\tau k H_I) \ \delta X_i^\parallel$, allows to further simplify the EOM to
\begin{equation}\label{eq:eom_longitudinal_tilde}
0 = \left[ \partial_\tau^2 + \frac{m_X^2 - 2 (\kappa + 1) H_I^2}{\tau^2 H_I^2} + k^2 \right] \widetilde{\delta X}^\parallel.
\end{equation}
This equation allows the Bunch-Davies vacuum as initial condition (cf.~\cite{Dimopoulos:2008yv})
\begin{equation}\label{eq:bunch_davies}
\widetilde{\delta X}^\parallel \xrightarrow[]{\tau \to -\infty} \frac{1}{\sqrt{2k}} \mathrm{e}^{-i k \tau}.
\end{equation}
Eq.~\eqref{eq:eom_longitudinal_tilde} with initial condition Eq.~\eqref{eq:bunch_davies} has the solution~\cite{thomasMaster}
\begin{equation} \label{eq:vector_long_before_horizon_exit_sol}
\delta X^\parallel \equiv -\frac{\tau k H_I}{|\widetilde{m}|} \widetilde{\delta X}^\parallel = -\frac{\tau k H_I}{|\widetilde{m}|} \sqrt{-\pi \tau} \frac{\mathrm{e}^{i \frac{\pi}{2} \left( \widetilde{\nu} + \frac{1}{2} \right)}}{1 - \mathrm{e}^{2 i \pi \widetilde{\nu}}} \left[ J_{\widetilde{\nu}} (-k \tau) - \mathrm{e}^{i \pi \widetilde{\nu}} J_{-\widetilde{\nu}} (-k \tau) \right],
\end{equation}
where $J_{\pm \widetilde{\nu}} (\cdot)$ are Bessel functions of the first kind and we introduce $\widetilde{\nu} \equiv 1/2 \sqrt{1 + 8 (\kappa + 1) - 4 m_X^2/H_I^2}$. This is the early time solution shown in Fig.~\ref{fig:numerical_interpolation}.

In the superhorizon limit, i.e. when $-\tau k  \ll 1$, Eq.~\eqref{eq:vector_eom_inflation} can be recast as
\begin{equation}
0 = \left[ \partial_\tau^2 - \frac{2 \tau k^2 H_I^2}{\widetilde{m}^2} \partial_\tau + \frac{\widetilde{m}^2}{\tau^2 H_I^2} + k^2 \right] \delta X^\parallel.
\end{equation}
This equation can be explicitly solved in terms of confluent hypergeometric functions of the first kind. By making use of relations $m_X^2 \ll 2 \kappa H_I^2$ and $-\tau k \ll 1$, the solution can be approximately expressed as
\begin{equation} \label{eq:vector_long_after_horizon_exit_sol}
\delta X^\parallel \approx 2^{-2 - \half\beta_-} (k \tau)^{-1+\half \beta_-} \left[ \widetilde{C}_1\, (k \tau)^{3-\beta_-} + \widetilde{C}_2\, 2^{3-\beta_-} \right] \approx \widetilde{C}_2\, 2^{1 - \frac{1}{2} \beta_-} (k \tau)^{-1 + \frac{1}{2} \beta_-},
\end{equation}
where $\widetilde{C}_1$ and $\widetilde{C}_2$ are constants that depend on the initial conditions. The second approximation holds long after horizon exit.
This the late time solution that is depicted in Fig.~\ref{fig:numerical_interpolation}.

The full match of Eqs.~\eqref{eq:vector_long_before_horizon_exit_sol} and~\eqref{eq:vector_long_after_horizon_exit_sol} cannot be performed analytically, but it is easy to see that the matching is only possible if the $k$-dependence of the coefficients is precisely $\widetilde{C}_{1,2} \sim 1/\sqrt{k}$. Using this knowledge, we can extract the momentum dependence and define $C_2^\prime$ such that
\begin{equation}
{\delta\chi}^\parallel \simeq C^\prime_2 2^{1-\half\beta_-} \frac{H_I}{k^{3/2}} \left( \frac{k}{aH_I} \right)^{\half\beta_-},
\end{equation}
in terms of scale factors rather than conformal time. This is precisely the result derived in Sec.~\ref{sec:isocurvature_vector}.


\bibliography{literature}


\end{document}